\documentclass[9pt,twocolumn,twoside]{optica}
\setboolean{shortarticle}{false}
\setboolean{minireview}{false}

\usepackage{graphicx}%
\usepackage{bm}%
\usepackage[amssymb]{SIunits}
\usepackage{float}
\usepackage{color}
\usepackage{amssymb}
\usepackage{wasysym}

\definecolor{commentColorWTJ}{rgb}{0.6,0.2,0.0}
\definecolor{commentColorASN}{rgb}{0.2,0.6,0.0}
\definecolor{commentColorRNP}{rgb}{0.0,0.0,1.0}
\definecolor{commentColorRVL}{rgb}{0.0,0.5,0.5}

\usepackage[normalem]{ulem}%

\newcommand{\omegac}{\omega_\text{c}}
\newcommand{\omegamu}{\omega_{\mu}}
\newcommand{\omegam}{\omega_{\textrm{m}}}

\newcommand{\opd}[2]{\mbox{$\hat{#1}_{#2}^{\dagger}$}}  %
\newcommand{\op}[2]{\mbox{$\hat{#1}_{#2}$}}

\newcommand{\kappae}{\kappa_{\textrm{e}}}
\newcommand{\kappai}{\kappa_{\textrm{i}}}
\newcommand{\kappamu}{\kappa_{\mu}}
\newcommand{\kappamue}{\kappa_{\mu,\textrm{e}}}
\newcommand{\kappaemu}{\kappa_{\textrm{e,}\mu}}
\newcommand{\gmu}{g_{\mu}}

\newcommand{\Qm}{Q_{\textrm{m}}}
\newcommand{\aout}{a_{\textrm{out}}}
\newcommand{\ain}{a_{\textrm{in}}}
\newcommand{\cin}{c_{\textrm{in}}}
\newcommand{\cout}{c_{\textrm{out}}}

\newcommand{\bin}{b_{\textrm{in}}}
\newcommand{\Aa}{A_{\textrm{a}}}
\newcommand{\Ab}{A_{\textrm{b}}}
\newcommand{\Ac}{A_{\textrm{c}}}
\newcommand{\etaab}{\eta_{\textrm{ab}}}
\newcommand{\etaba}{\eta_{\textrm{ba}}}
\newcommand{\etacb}{\eta_{\textrm{cb}}}
\newcommand{\etabc}{\eta_{\textrm{bc}}}
\newcommand{\etabcb}{\eta_{\textrm{bcb}}}
\newcommand{\etaaba}{\eta_{\textrm{aba}}}
\newcommand{\etaabc}{\eta_{\textrm{abc}}}
\newcommand{\etacba}{\eta_{\textrm{cba}}}
\newcommand{\Cab}{C_{\textrm{ab}}}
\newcommand{\Cbc}{C_{\textrm{bc}}}

\newcommand{\gammai}{\gamma_{\textrm{i}}}

\newcommand{\gammaOM}{\gamma_{\textrm{OM}}}
\newcommand{\gammatot}{\gamma_{\textrm{tot}}}
\newcommand{\gammae}{\gamma_{\textrm{e}}}
 
\newcommand{\nc}{n_{\textrm{c}}}
\newcommand{\ncsty}{\bar{n}_{\textrm{c}}}

\newcommand{\Pin}{P_{\textrm{in}}}
\newcommand{\Gammatot}{\Gamma_{\textrm{tot}}}

\title{Lithium Niobate Piezo-optomechanical Crystals}

\author[1]{Wentao Jiang}
\author[1]{Rishi N. Patel}
\author[1]{Felix M. Mayor}
\author[1]{Timothy P. McKenna}
\author[1]{Patricio Arrangoiz-Arriola}
\author[1]{Christopher J. Sarabalis}
\author[1]{Jeremy D. Witmer}
\author[1]{Rapha\"el Van Laer}
\author[1]{Amir H. Safavi-Naeini}%

\affil[1]{Department of Applied Physics and Ginzton Laboratory, Stanford University, 348 Via Pueblo Mall, Stanford, California 94305, USA}

\begin{abstract}
Demonstrating a device that efficiently connects light, motion, and microwaves is an outstanding challenge in classical and quantum photonics. We make significant progress in this direction by demonstrating a photonic crystal resonator on thin-film lithium niobate (LN) that simultaneously supports high-$Q$ optical and mechanical modes, and where the mechanical modes are coupled piezoelectrically to microwaves. For optomechanical coupling, we leverage the photoelastic effect in LN by optimizing the device parameters to realize coupling rates $g_0/2\pi\approx 120~\textrm{kHz}$. An optomechanical cooperativity $C>1$ is achieved leading to phonon lasing. Electrodes on the nanoresonator piezoelectrically drive mechanical waves on the beam that are then read out optically allowing direct observation of the phononic bandgap. Quantum coupling efficiency of $\eta\approx10^{-8}$ from the input microwave port to the localized mechanical resonance is measured. Improvements of the microwave circuit and electrode geometry can increase this efficiency and bring integrated ultra-low-power modulators and quantum microwave-to-optical converters closer to reality.
\end{abstract}

\setboolean{displaycopyright}{false}

\begin{document}

\maketitle

\section{Introduction}

Optomechanical crystals (OMC) provide a powerful platform for engineering interactions between photons and phonons. Co-localizing optical and mechanical modes allows light to control and readout mechanical motion, leading to a large variety of quantum optomechanical experiments~\cite{chan2011laser,PhysRevLett.108.033602,hill2012coherent, cohen2015phonon,marinkovic2018optomechanical} and enabling routing and transduction of classical and quantum signals between microwave-frequency phonons and optical photons~\cite{PhysRevLett.105.220501,fang2016optical,fang2017generalized,patel2018single,riedinger2018remote}. Silicon optomechanical crystals have strong coupling rates but lack intrinsic piezoelectricity \cite{chan2012optimized}. Capacitive and electrostrictive forces in thin-film silicon have been used to drive mechanical resonances~\cite{Weinstein2010,VanLaer2018,Kalaee2019}, but a viable way of efficiently coupling to the high-frequency wavelength-scale mechanical modes of an optomechanical crystal resonator is yet to be demonstrated. Thus approaches to coupling OMCs to microwave photons and superconducting qubits have focused on piezoelectric materials such as aluminum nitride (AlN)~\cite{pernice2012high,fan2013aluminum,bochmann2013nanomechanical,vainsencher2016bi}, gallium arsenide (GaAs)~\cite{ding2011wavelength,balram2014moving,balram2016coherent,forsch2018microwave} and gallium phosphide (GaP)~\cite{mitchell2014cavity,schneider2017optomechanics}. However, no platform to date has succeeded in simultaneously demonstrating high optical and mechanical quality factors, large optomechanical coupling, and good piezoelectric coupling efficiency. Recent work on nanopatterned lithium niobate (LN) has addressed several of these weaknesses, by showing that high mechanical $Q$ and efficient piezoelectric transduction of localized modes are possible~\cite{Arrangoiz-Arriola2018,shao2019high}, while excellent optical $Q$'s have been demonstrated in photonic crystals~\cite{Liang2017, li2018high} and microring structures~\cite{wang2018ultrahigh}. In this work we demonstrate an LN device with sufficiently large optical and mechanical $Q$s and optomechanical coupling to realize efficient optical transduction of microwave frequency phonons. Considering that efficient microwave transduction has already been realized in the same material system with phononic crystal devices~\cite{Arrangoiz-Arriola2018}, our result solves a key remaining technical challenge required for an LN quantum microwave-to-optical transducer. 

Here we demonstrate one-dimensional (1D) nanobeam OMCs on lithium niobate (LN) with optical quality factors above 300,000 at a wavelength of $\lambda\approx1550~\text{nm}$, and a mechanical mode frequency close to $ 2~\textrm{GHz}$ possessing quality factors as high as 17,000 at 4 kelvin. The high frequency mechanical mode and narrow optical linewidth allows our system to operate in the resolved-sideband regime, where the mechanical frequency greatly exceeds the cavity decay rate. An important parameter of the system is the zero-point coupling rate $g_0$ which characterizes the optical frequency jitter induced by the zero-point motion of the cavity. We make calibrated measurements of optomechanical backaction in addition to measurements of electromagnetically induced transparency (EIT) and absorption (EIA)  with different optical pump powers and detunings to extract a zero-point coupling rate of $g_{0}/2\pi \approx 120~\textrm{kHz}$. The wide bandgap ($\sim3.7~\textrm{eV}$) of LN allows us to pump $n_\text{pump}\sim 1.5\times 10^{5}$ intracavity photons without significant degradation of the optical linewidth. This enables us to realize sufficiently large cooperativity $C\propto n_\text{pump}$ to observe mechanical lasing under the action of a blue-detuned optical pump. Phonon lasing is an important signature of optomechanical cooperativity $C > 1$, which occurs when the optomechanical backaction exceeds the intrinsic mechanical dissipation in the system, a requirement for realizing efficient phonon-photon conversion~\cite{safavi2011proposal,Aspelmeyer2014,Safavi-Naeini2019,kippenberg2005analysis,jiang2016chip}. Our system performs favorably when compared to other piezo-optomechanical approaches. In particular, OMCs on GaAs show $g_{0}/2\pi$ on the order of $1~\textrm{MHz}$~\cite{balram2014moving}, but have not operated in the sideband-resolved regime due to low optical $Q\sim40,000$. Moreover, $g_{0}/2\pi\sim 100~\textrm{kHz}$ has been achieved on AlN~\cite{vainsencher2016bi}, but our LN device possesses somewhat higher optical and mechanical quality factors.  Finally we note that all three platforms, AlN~\cite{o2010quantum}, GaAs~\cite{Sletten2019resolving}, and LN~\cite{Arrangoiz-Arriola2018,arrangoiz2019resolving} have demonstrated compatibility with superconducting qubit technology and form promising platforms for quantum transduction.

The piezoelectric property of LN enables us to electrically drive the localized mechanical mode with an oscillating microwave electric field using fabricated on-chip electrodes. The electrically generated phonons are detected optomechanically allowing us to measure the piezo-optomechanical response of the OMC. More generally, electrical driving allows us to probe the full mechanical spectrum of the device with far greater sensitivity than is achieved by measuring thermal Brownian motion, as it allows us to drive the modes to a larger amplitude than is possible thermally. This leads to a remarkably clear signature of the mechanical bandgap of the nanobeam mirror unit cell that is visible from the piezo-optomechanical response. Furthermore, we analyze the piezo-optomechanical coupling by calibrating the coherent phonon number from the electrical drive against the known thermal phonon occupancy to infer the effective microwave-mechanics coupling rate $g_{\mu}/2\pi \sim 2~\textrm{kHz}$ that characterizes the rate at which microwave frequency phonons would interact with a microwave frequency resonator (assuming a $50~\Omega$ resonator impedance). Comparing to the mechanical losses,  this coupling is about two orders of magnitude away from achieving sufficiently large cooperativity to enable coherent microwave-to-optical conversion.

We have demonstrated that the LN on silicon platform can be engineered to support strong optomechanical coupling between telecom photons and GHz phonons. This further demonstrates that lithium niobate is a material system with enormous potential for quantum electro-optic and optomechanical transduction~\cite{witmer2017high,wang2018nanophotonic,wang2018integrated,bochmann2013nanomechanical,arrangoiz2016engineering,balram2016coherent,Arrangoiz-Arriola2018}. In addition to piezo-optomechanical transduction, the electro-optic and nonlinear optical properties of LN enable active integrated optical elements and enhanced optical parametric processes~\cite{wang2017second,wang2018nanophotonic,wang2018integrated,wang2018ultrahigh,jiang2018nonlinear}.

\begin{figure*}[t]
\centering
\includegraphics[scale=0.6]{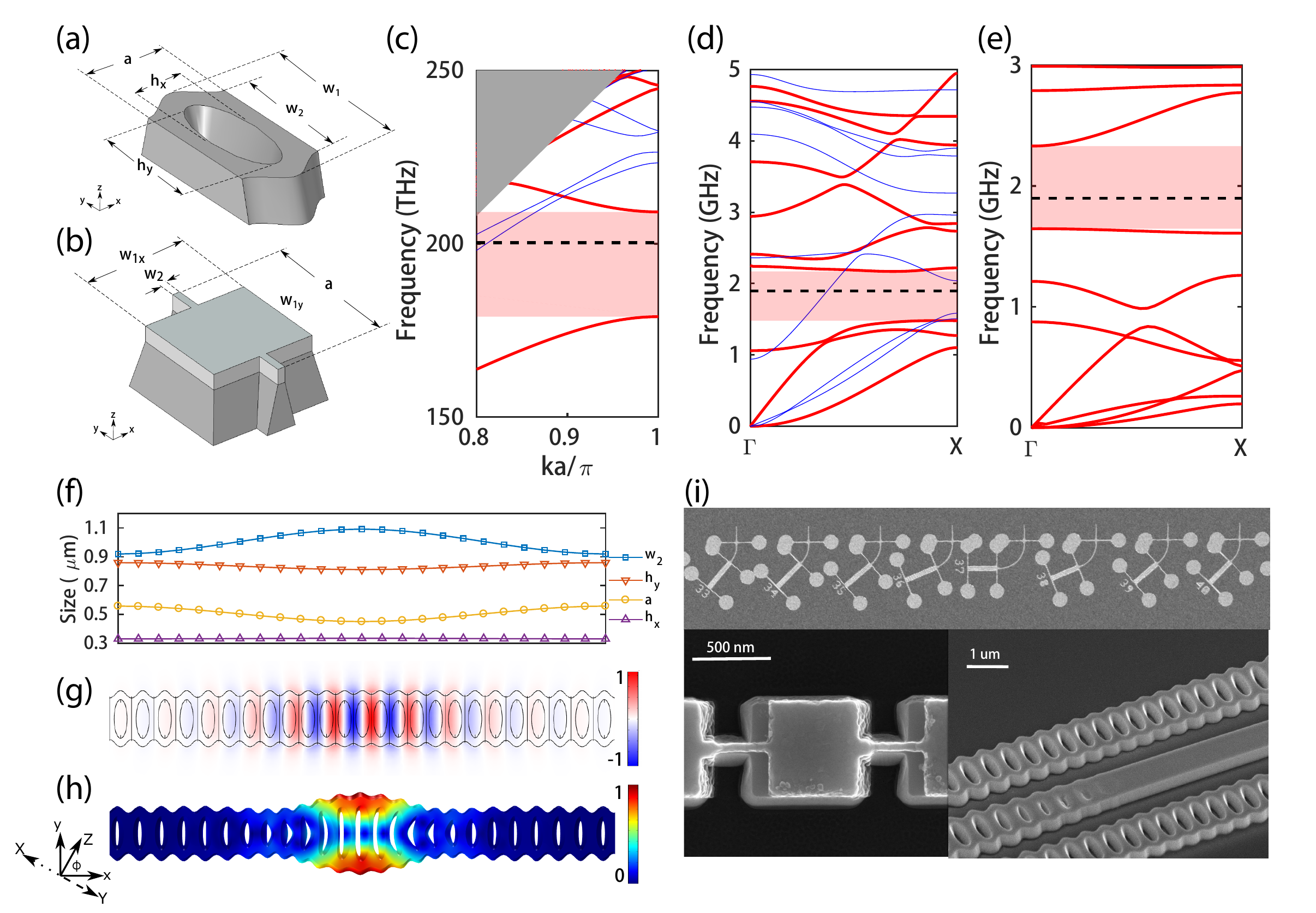}
\caption{\label{fig1:design} One-dimensional optomechanical crystal (OMC) design. (a) Unit cell geometry of the nanobeam OMC, showing definitions of the parameters. (b) Unit cell geometry of the 1D phononic shield. The direction of the periodicity is perpendicular to the direction of the nanobeam. The (c) optical and (d) mechanical band structure of the nanobeam unit cell. The bands are classified by y-symmetry (red) and y-antisymmetry (blue). In (c), the gray shaded region represents the continuum of radiation modes. The pink shaded regions highlight the bandgap for quasi-TE optical modes (c) and symmetric mechanical modes (d). The dashed black lines correspond to the localized fundamental TE optical mode and the mechanical breathing mode of the nanobeam respectively. (e) Mechanical band structure of the 1D phononic shield unit cell, showing a complete bandgap around 2 GHz. (f) Variations of unit cell parameters along the nanobeam. $w_{1}$ is kept constant for all unit cells. (g) $E_{y}$ component of the optical mode. (h) Displacement field of the breathing mode. The global and material coordinates are shown (see text for description). (i) Scanning electron micrograph (SEM) of the fabricated device. Top: one row of devices with different nanobeam orientations, taken before the aluminum lift-off step. Bottom left: top view of the 1D phononic shield region, showing both the LN pattern and the Al electrodes. Misalignment on the order of $20~\nano\meter$ is visible. Bottom right: SEM of two nanobeams coupled to one reflector.} 
\end{figure*}

\section{Design}

We begin with the design and simulation of the 1D OMC on LN. Due to imperfections in the LN nanofabrication process, the resulting inside and outside sidewall angles are approximately $\theta_{\textrm{in}} = 22^\circ$ and $\theta_{\textrm{out}} = 11^\circ$ from the vertical direction. As a result, the nanobeam no longer possesses the reflection symmetry about the $xy$-plane, which we call $z$-symmetry. Given a structure that has $y$-symmetry, as long as the material properties also remain invariant under reflection about the $xz$-plane, the modes can be classified according to their $y$-symmetry. For optical waves, material properties will be invariant as long as the $Z$ is in the longitudinal direction of the beam. For mechanical waves in LN, the situation is a bit more complicated due to the anisotropic elastic properties that generally break the $y$-symmetry, even for a symmetric structure with the $Z$-axis aligned along the nanobeam. In the special case where the nanobeam is fabricated on $Y$-cut LN and is parallel to the crystal $Z$ axis, one of the crystal mirror planes coincides with the $y$-symmetry plane of the nanobeam geometry, and the $y$-symmetry is restored. Under this configuration, the optical and mechanical modes can be classified according to their $y$-symmetry properties.

We adapt the design recipe demonstrated for silicon OMCs~\cite{chan2012optimized} to LN considering our etch properties and material anisotropy. We first design a mirror unit cell with quasi-TE optical bandgap of $30~\textrm{THz}$ at $X$-point around $194~\textrm{THz} $ and a partial mechanical bandgap from $1.48~\textrm{GHz}$ to $2.17~\textrm{GHz} $ for $y$-symmetric mechanical modes. The geometry of the mirror cell is shown in fig.~\ref{fig1:design}(a), with outer width $w_{1} = 1247~\nano\meter $, inner width $w_{2} = 919~\nano\meter $, $h_{x} = 330~\nano\meter $, $h_{y} = 859~\nano\meter$, thickness $t = 300~\nano\meter $ and lattice constant $a = 558~\nano\meter$. We use a sinusoid-shaped outer edge to realize a larger air-hole. Fig.~\ref{fig1:design}(c) and fig.~\ref{fig1:design}(d) show the optical and mechanical band structures of the mirror cell. The corresponding bandgaps are emphasized as pink shaded regions.

To localize optical and mechanical modes, we focus on the $X$-point fundamental quasi-TE optical mode and the $\Gamma$-point mechanical breathing mode. By reducing the lattice constant and the separation between the air-holes, the $X$-point optical mode shifts up in frequency and the $\Gamma$-point breathing mode frequency is reduced, putting both modes inside their corresponding bandgaps.  Using multivariable genetic optimization, defect cell geometry parameters are found to be $(a, w_{2}, h_{x}, h_{y}) = (450, 1092, 334, 811 )~\nano\meter$ by maximizing $g_{0}\min(Q,Q_{\textrm{th}})/Q_{\textrm{th}}$, where $g_{0}$ is the optomechanical coupling rate, $Q$ is the radiation-limited optical quality factor and $Q_{\textrm{th}} = 10^{7} $ is chosen to ignore unachievably high $Q$~\cite{chan2012optimized}. The full nanobeam is generated by a smooth transition from the mirror cell geometry to the defect cell geometry. The variation of the parameters is shown in fig.~\ref{fig1:design}(f) together with the finite-element method simulated~\cite{COMSOL} fundamental TE optical mode (fig.~\ref{fig1:design}(g)) and mechanical breathing mode (fig.~\ref{fig1:design}(h)). The LN thickness $t$ and the outer edge $w_{1}$ are kept the same for the full nanobeam. The global and material coordinate systems are shown at the bottom-left of fig.~\ref{fig1:design}(h). The in-plane rotation angle $\phi$ is defined as the angle between crystal axis $Z$ and global axis $x$ for both $X$-cut and $Y$-cut LN (supplementary material). The dashed (dotted) arrow represents crystal $Y$($X$) axis for $X$($Y$)-cut LN. We obtained a radiation-limited optical quality factor $Q\sim 4\times 10^{6}$ and an effective mode volume as small as $\sim 0.2 (\lambda/n)^{3} $ for the simulated fundamental optical mode at $\omegac/2\pi = 200~\textrm{THz}$. The simulated mechanical breathing mode is at $\omegam/2\pi = 2.1~\textrm{GHz}$ with effective mass $m_{\textrm{eff}} = 645~\textrm{fg}$ and zero-point motion $x_{\textrm{zp}} = 2.48~\textrm{fm}$. The mode profiles are modified slightly by the crystal orientation. The primary effect of the orientation however is the variation in the optomechanical coupling due to anisotropic nature of the photoelastic effect -- this is studied in more detail in the supplementary material. 

The lack of $y$-symmetry in general induces leakage of phonons into other propagating modes (blue bands in Fig.~\ref{fig1:design}d) of the beam by making nominally disallowed transitions possible as has been studied in previous work on silicon structures~\cite{patel2017engineering}. To reduce the effect of this source of mechanical dissipation, we implement a 1D phononic shield (PS) to anchor the nanobeam~\cite{safavi2010design,chan2012optimized}. The geometry of the PS unit cell is shown in fig.~\ref{fig1:design}(b), including the top aluminum electrode layer with thickness $t_{\textrm{m}}=100~\nano\meter$. The dimensions of this PS unit cell are chosen to be $(a, w_{1\textrm{x}}, w_{1\textrm{y}}, w_{2}) = (1000, 600, 650, 50)~\nano\meter $ such that the mechanical band structure of the unit cell opens up a full bandgap from $1.65~\textrm{GHz}$ to $2.33~\textrm{GHz}$ (fig.~\ref{fig1:design}(e)). The effect of the metallization layer has been taken into account in these simulations.

\begin{figure*}[t]
\centering
\includegraphics[scale=0.51]{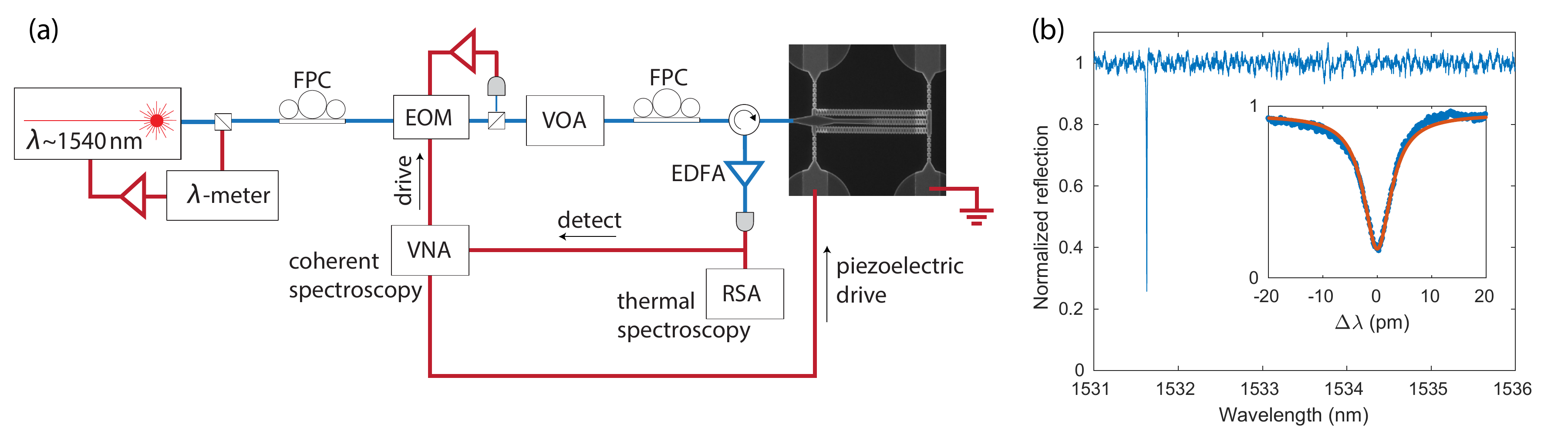} 
\caption{\label{fig2:setup-optics} Measurement setup and optical mode characterization. (a) Simplified diagram of the measurement setup. Two LN optomechanical crystals are side-coupled to a reflector. The reflection spectrum is recorded for optical characterization. The thermal mechanical motion of the nanobeam encoded in the optical noise power spectrum is measured by the realtime spectrum analyzer (RSA). A weak optical side band is generated with the vector network analyzer (VNA) and electro-optic modulator (EOM) for coherent spectroscopy. The mechanical motion can be also piezoelectrically driven and the transduced optical sidebands are measured by the high-speed photodetector and the VNA. (b) Optical reflection spectrum of an LN OMC. A zoomed-in wavelength sweep shows a loaded optical quality factor $Q = 2.5\times 10^{5}$ and corresponding intrinsic quality factor $Q_{i} = 3.5\times 10^{5}$ (inset, blue: data, red: fit). Variable optical attenuator (VOA), erbium-doped fiber amplifier (EDFA), fiber polarization controller (FPC).}
\end{figure*}

\section{Fabrication}

In contrast to silicon, nanofabrication of LN is known to suffer from a lack of an effective chemical etch. However, since the development and commercial availability of thin-film LN on silicon oxide~\cite{hu2009lithium,poberaj2012lithium}, high-$Q$ micro-ring resonators and photonic crystals have been demonstrated with physical argon milling and reactive ion etching ~\cite{wang2014integrated,hartung2008fabrication,zhang2017monolithic,krasnokutska2018ultra,Liang2017,li2018high}. We adopt an argon milling process similar to Ref.~\cite{wang2014integrated} for patterning the OMC structure.

The OMCs are fabricated on a lithium niobate on silicon (LNOS) wafer, with an initial LN thickness of $t\approx 500~\nano\meter$. The LN layer is first thinned to $ t = 300~\nano\meter $ with argon ion milling. The nanobeam geometry is defined by electron beam lithography (EBL) with HSQ resist. The pattern is then transferred to the LN layer by argon ion milling. The remaining HSQ is removed with 2 minute 5:1 buffered oxide etch (BOE). A second aligned EBL with CSAR resist followed by a liftoff process defines the aluminum metal layer. The device is finally released with a masked $\textrm{XeF}_{\textrm{2}}$ dry etch that selectively removes the silicon. Electrical contacts between on-chip electrodes and aluminum wires are made with ultrasonic wire-bonding. Tapered optical mode converters are patterned on the LN layer with efficiency $\eta \sim 50\%$ for coupling the OMCs to a lensed fiber~\cite{patel2018single}.

Fig.~\ref{fig1:design}(i) shows scanning electron micrographs of the resulting structures taken before the release step. A notable feature of the sample is that it contains many devices with the identical OMC design, but where we have rotated the orientation of the nanobeam with respect to the crystal axis (see also Ref.~\cite{balram2014moving} for similar measurements on GaAs). This is done to give us a better understanding of the anisotropic photoelastic property of LN and its effect on the performance of the OMCs.

\section{Optomechanical Characterization and Mechanical Lasing}

\begin{figure*}[t]
\centering
\includegraphics[scale=0.4]{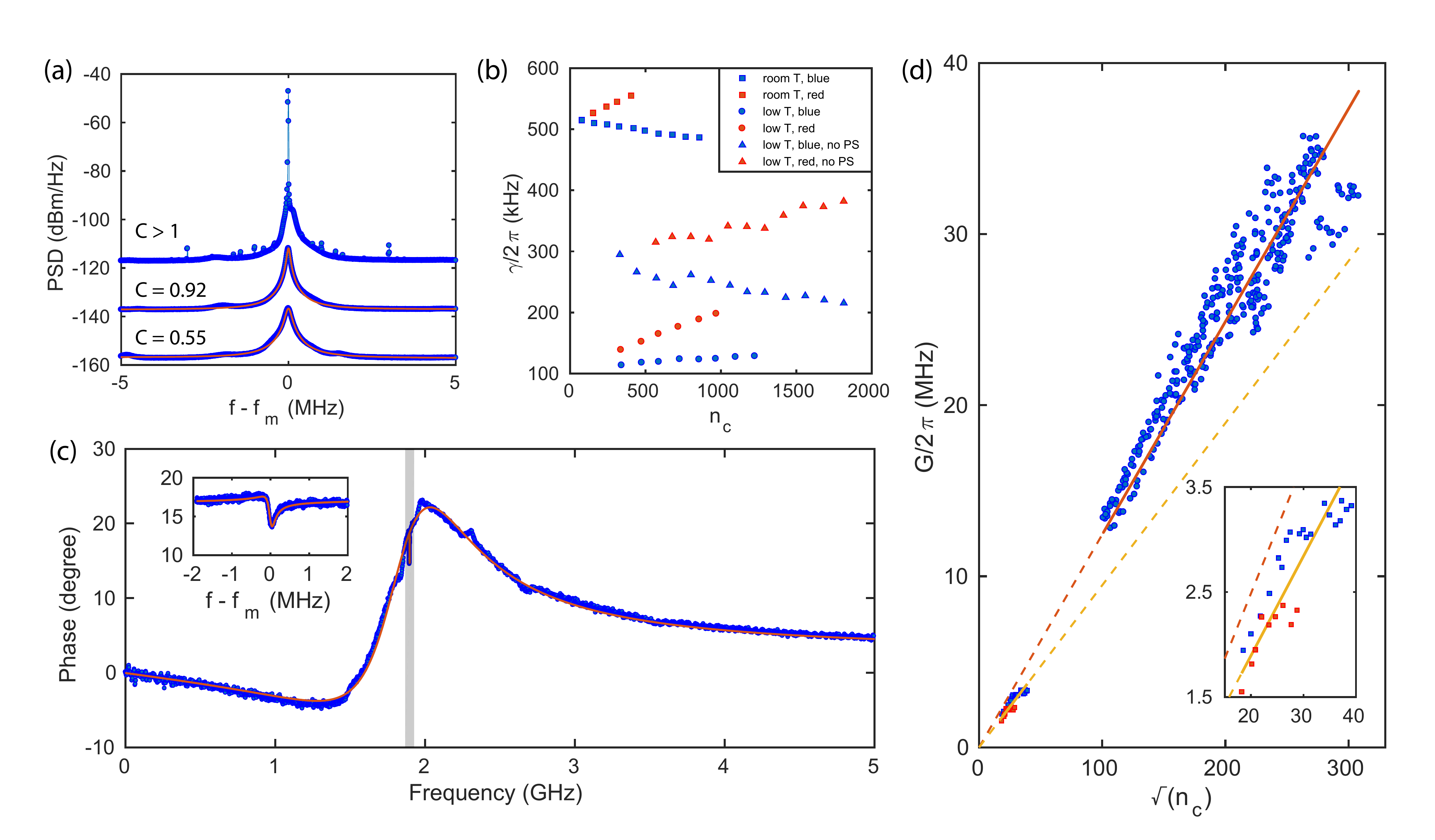}
\caption{\label{fig3:EITdata} EIT and optomechanical backaction measurements. (a) Power spectral density of the mechanical motion induced optical sideband. Measured spectrum with same optical power and different detunings are shown with corresponding cooperativities, vertically displaced for viewing purposes. When a pump laser beam is blue detuned near $\Delta \approx -\omegam$, mechanical lasing results as the cooperativity exceeds unity. The spectrum without mechanical lasing is fitted (red) for the total mechanical linewidth $\gammatot$. (b) Total mechanical linewidth versus intracavity photon numbers under different conditions (blue: $\Delta = -\omegam$, red: $\Delta = \omegam$). For the low temperature results, one OMC without the 1D phononic shield are measured for comparison ($\triangle$). (c) Measured phase response of the optical probe beam at low power pump with detuning $\Delta = \omegam$ at low temperature (blue dots) and fit (red line). Inset: zoom-in data and fit near the transparency window. (d) Effective optomechanical coupling rate $G$ extracted from the EIT responses at low temperature versus the square root of intracavity photon numbers $\sqrt{\nc} $. High-power optical pump at detunings $ -\omegam < \Delta < 0$ is adopted to achieve high $\nc$ without introducing mechanical lasing. Linear fits for high $\nc$ (orange) and low $\nc$ (yellow) measurements are both shown for comparison. Inset: detail of the low $\nc$ region, with data from both blue (blue) and red (red) detuned pump. }
\end{figure*}

We measured the devices at both room temperature and at low temperature, by cooling to $4~\text{K}$ inside a closed-cycle Montana Instruments cryostat.  The actual temperature of the phonon mode in the cryostat is closer to $\sim 20~\textrm{K}$. This is determined by calibrated thermal spectroscopy as explained in Ref.~\cite{patel2018single}. The measurement setup, shown in fig.~\ref{fig2:setup-optics}(a), is suited for understanding the optical and mechanical properties of the system. The optical properties were first characterized by scanning a continuous-wave tunable laser and monitoring the reflection spectrum. Fig.~\ref{fig2:setup-optics}(b) shows a typical optical reflection spectrum with one sharp resonance from the localized optical mode. A narrow scan near the resonance is fit to a lorentzian lineshape to determine the linewidth and quality factor (fig.~\ref{fig2:setup-optics}(b) inset). For the device under consideration, a loaded $Q$ of $2.5\times 10^{5}$ and an intrinsic $Q$ of $3.5\times 10^{5}$ are extracted, corresponding to a total linewidth $\kappa/2\pi = 776~\textrm{MHz}$ that is significantly smaller than the mechanical frequency and puts us in the resolved sideband regime of cavity optomechanics~\cite{Aspelmeyer2014}. Similar measurements taken over multiple devices, as detailed in the supplementary materials, show no apparent correlation between the optical quality factors and the crystal orientations. 

A well known effect in optical resonators is the shift in their frequency proportional to intracavity photon number $\nc$. This nonlinearity can result from the thermo-optic, or Kerr effect. In our measurements, we observe a strikingly different relation between intracavity photon number and cavity resonance shift. As $\nc $ is increased, we see that the cavity frequency shifts by a frequency proportional to $\nc^{2}$. The shift becomes easily measurable at roughly $\nc \gtrsim 10^{4}$, when it exceeds the linewidth. At these values and larger, it is observed to be quadratic. In addition to measuring the frequency shift, we use a sideband probe to measure the cavity linewidth at different optical powers and detunings (see SI for full dataset). Based on the amount of linewidth broadening that is observed ($\sim 15\%$ at $\nc=1.5\times10^5$), we attribute the quadratic thermal shift to absorption of the intracavity photons, where this absorption rate itself has a linear dependence on $\nc$.  This dependence distinguishes it from  green-induced infrared absorption (GRIIRA)~\cite{furukawa2001green}, in which case the absorption rate of intracavity photons would scale as $\nc^2$ (the second harmonic intensity) instead of the observed $\nc$. Moreover, it is likely not due to absorption of second-harmonic generated photons, which would produce the correct $\nc^2$ thermal shift, but would also require a significant reduction in the cavity linewidth to explain the observed thermal shift. A detailed understanding of the physical mechanism behind this heating and the development of methods for its mitigation (such as surface treatment~\cite{forsch2018microwave} or Mg-doping~\cite{furukawa2001green}) will be the subject of future investigations.

The optomechanical coupling between the localized optical mode and mechanical mode gives rise to an interaction Hamiltonian $ \op{H}{\textrm{int}} = \hbar g_{0} \opd{a}{}\op{a}{}(\op{b}{}+\opd{b}{}) $, where $g_{0}$ is the zero-point optomechanical coupling rate, $\op{a}{}$ and $\op{b}{}$ are the annihilation operator for the optical and mechanical modes respectively. When the optical mode is driven by a laser beam up to an intracavity photon number $\nc\gg 1$, the interaction can be linearized to $ \op{H}{\textrm{int}} = \hbar G (\op{a}{}+\opd{a}{})(\op{b}{}+\opd{b}{}) $, and $G = g_{0}\sqrt{\nc}$ is the effective optomechanical coupling rate~\cite{Aspelmeyer2014}. The optomechanical (OM) backaction leads to a frequency shift and an extra damping rate of the mechanical mode given by~\cite{Aspelmeyer2014}
\begin{equation}
    \gammaOM = G^{2} \left( \frac{\kappa}{\kappa^2/4 + (\Delta-\omegam)^2} -  \frac{\kappa}{\kappa^2/4 + (\Delta+\omegam)^2}  \right),
\end{equation}
where $\kappa$ is the total  optical mode linewidth, $\Delta = \omega_{\textrm{c}} - \omega_{\textrm{l}}$ is the cavity-laser detuning and $\omegam$ is the mechanical mode frequency. In the sideband-resolved regime where $\omegam > \kappa$, only the first term is appreciable for a red detuned pump laser with $\Delta \sim \omegam$, while for blue side pumping, the second term dominates. As shown in fig.~\ref{fig2:setup-optics}(a), the mechanical properties of the system can be characterized by either measuring the transduced mechanical thermal motion in the optical noise power spectral density (thermal spectroscopy), or by measuring the effect of the mechanical response on a weak optical probe generated by the electro-optic modulator (coherent spectroscopy).

The optomechanical coupling rates are observed to peak at $\phi = 135^\circ$ on $X$-cut LN. This is in agreement with simulations outlined in detail in the supplementary materials. We focus on one device with this specific orientation and characterize its mechanical and optomechanical properties. We first perform the thermal spectroscopy at low temperature with a blue detuned laser pump, where the OM backaction is simplified to
\begin{equation}
\gammaOM = -\frac{G^{2}\kappa}{\kappa^{2}/4+(\Delta + \omegam)^{2}}.
\end{equation}
We fix the pump power $\Pin \sim 0.2~\textrm{mW}$ and step the detuning towards $\Delta = -\omegam$ to increase the optomechanical cooperativity $ C \equiv |\gammaOM/\gammai|$, where $\gammai$ is the intrinsic mechanical linewidth.  Fig.~\ref{fig3:EITdata}(a) shows three different power spectral densities of the reflected pump laser at different values of cooperativity. When $C > 1$ is reached, the total mechanical linewidth $\gammatot = \gammai + \gammaOM$ becomes negative. The system enters an oscillating regime and a mechanical lasing peak emerges in the power spectrum.

To extract the intrinsic mechanical linewidth $\gammai$ and zero-point optomechanical coupling rate $g_{0}$ of the same device, we fit $\gammatot $ from the power spectrum. We fix cavity-laser detuning $\Delta = \pm \omegam $ so the OM backaction is given by $\gammaOM = \pm 4G^{2}/\kappa = \pm 4 g_{0}^{2} \nc/\kappa $, resulting in a linear relationship between $\gammaOM$ and $\nc$. We adopted small pump powers $\Pin \sim 10~\micro\textrm{W}$ in order to access stable red-detuned measurements in the presence of a thermal cavity red shift. As shown in fig.~\ref{fig3:EITdata}(b), we measured the device at both room and low temperature and on the blue and red sides. The intrinsic mechanical loss $\gammai$ is reduced by a factor of $5$ after cooling the device. We observe that $\gammatot$ deviates from simple predictions only considering OM backaction, showing an asymmetry with respect to the intrinsic linewidth between blue and red detuned laser pump frequency. The deviation becomes larger when the device is cooled down, where no linewidth reduction is observed with a blue detuned pump laser. We attribute the deviation to thermally-induced linewidth broadening which scales linearly with respect to $\nc$ for low pump powers and is independent of the sign of the detuning. At higher pump powers, the thermally-induced broadening saturates near $\gammatot/\pi \approx 300~\textrm{kHz}$ and as a result, the OM backaction eventually dominates the linewidth change, leading to mechanical lasing. We combine the measurements from $\Delta = \pm \omegam $ at low powers to eliminate the thermal broadening. We measure $g_{0}/2\pi = 128~\textrm{kHz}$, intrinsic mechanical linewidth $\gammai/2\pi = 513~\textrm{kHz}$  at room temperature and $g_{0}/2\pi = 92~\textrm{kHz}$, $\gammai/2\pi = 109~\textrm{kHz}$  at low temperature. The cause of the temperature dependence of $g_0$ may be attributed to a temperature dependence of the photoelastic coefficients, but is not well undertood at this time. The corresponding mechanical quality factor is $\Qm \sim 3500$ at room temperature and $\Qm \sim 17000$ at low temperature. A  device with identical design but without a 1D phononic shield is also measured at low temperature for comparison ($\triangle$), showing $g_{0}/2\pi = 108~\textrm{kHz}$ and $\gammai /2\pi = 284~\textrm{kHz} $.

The optomechanical properties of the same device are also measured at low temperature using the optical sideband probe measurement (coherent spectroscopy). A typical frequency-dependent phase response of the probe is plotted in fig.~\ref{fig3:EITdata}(c). The response from the optical cavity is apparent in this measurement. Additionally, we observe electromagnetically induced transparency, visible in the phase variation near the mechanical frequency~\cite{weis2010optomechanically,safavi2011electromagnetically,massel2012multimode,safavi2014two}. The data shown was measured with optical pump detuning $ \Delta = \omegam $. The corresponding cooperativity $ C $ determines the size of the EIT feature. By fitting the phase response of the probe, we obtain an independent measurement of the optical mode linewidth $\kappa$, the intrinsic mechanical mode frequency $\omegam$ and linewidth $\gammai$, and also the effective optomechanical coupling strength $G$~\cite{safavi2011electromagnetically}.

\begin{figure*}[t]
\centering
\includegraphics[scale=0.68]{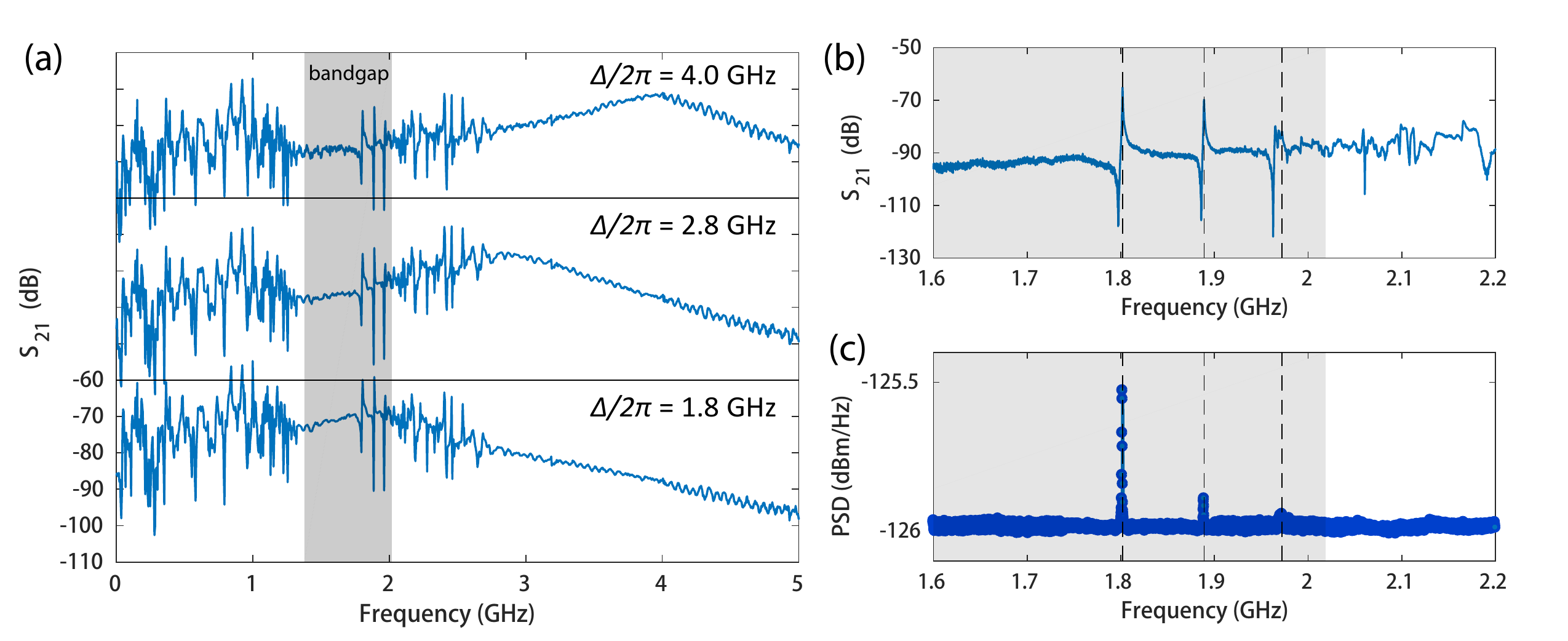}
\caption{\label{fig4:PiezoDrive} Piezo-optomechanical measurements. (a) Optical readout of electro-optically and piezoelectrically driven optical sideband for different pump detunings. Broad electro-optic effect induced peak shifts when changing pump detuning, while mechanical responses stay at the same frequency. The gray shaded area corresponds to the simulated mechanical bandgap of the nanobeam mirror unit cell. The mechanical response outside the bandgap is disordered but repeatable. (b) Piezo-optomechanical response near the upper bandgap edge. Peaks of the response match well with the thermal mechanical peaks from the power spectral density (c), corresponding to the localized mechanical breathing modes of the nanobeam.}

\end{figure*}

We extracted $G$ with different intracavity photon number $\nc$ in fig.~\ref{fig3:EITdata}(d). Measurements with $\nc$ approaching $10^{5}$ were achieved by using high pump powers at detunings $-\omegam < \Delta < 0$ to avoid mechanical lasing. The relation between $G$ and $\sqrt{\nc}$ is determine by fitting the low-power and high-power measurements independently. The results of these fits are shown in fig.~\ref{fig3:EITdata}(d) as yellow and orange lines, extended for comparison. From these fits, we extracted zero point couplings of $g_{0}/2\pi = 127 \pm 9~\textrm{kHz}$ for high $\nc$ and $g_{0}/2\pi = 96 \pm 10\textrm{kHz}$ for low $\nc$. We find reasonable agreement of $g_{0}$ between backaction measurement at low (high) temperature and EIT/EIA measurement at low (high) $\nc$. An important figure of merit is $g_0/\kappa$ which for this system is $10^{-4}$, making it comparable to competing implementations in AlN, SiN\textsubscript{x} and GaAs~\cite{Safavi-Naeini2019}.

\section{Piezo-optomechanical characterization}

We have fully characterized the optomechanical system by thermal and coherent spectroscopy. Beyond the conventional optomechanical measurements, the piezoelectricity of LN allows us to directly drive the mechanical motion of the OMC, revealing more details of the mechanical spectrum. To access piezoelectricity, we pattern aluminum electrodes on the LN layer that can be driven with an oscillating voltage. The separation between the edges of the electrodes are chosen to be $d_{\textrm{metal}}=22~\micro\meter$ such that there's no degradation on the optical quality factor. When a voltage is applied on the aluminum electrodes in the nanobeam mirror cell region, an electric field parallel to the nanobeam is generated and deforms the optomechanical crystal. The deformation can be sensitively read out by the laser utilizing the optomechanical coupling. 

The piezoelectric coupling between the OMC mechanical mode and the microwave channel (with impedance $Z_{0}=50~\Omega$) is modeled as an external coupling induced decay rate $ \gammae $. Our goal is to determine $\gammae$ by an optomechanically calibrated measurement of the piezomechanical driving efficiency. We pump the optical mode at $\Delta \sim \omegam  $ and drive the mechanics through the microwave channel. The coherent amplitude of the mechanical motion from standard input-output theory is (supplementary material)
\begin{equation}
    \beta = \frac{-\sqrt{\gammae} \cin}{i(\omegam - \omega_{\mu}) + \gammatot/2  },
\end{equation}
where $\cin$ is the input microwave field ($|\cin|^2$ is the microwave photon flux), $\omega_{\mu} \sim \omegam$ is the microwave signal frequency and $\gammatot=  \gammai + \gammae + \gammaOM$ is the total mechanical linewidth. The transduced optical field amplitude is
\begin{equation}
      \alpha =  \frac{ -i G\beta }{ i (\Delta - \omega_{\mu} ) + \kappa/2 }.
\end{equation}
We readout the beat tone between $\alpha$ and the pump laser, giving us the piezo-optomechanical $S_{21}(\omega)$.

Fig.~\ref{fig4:PiezoDrive}(a) shows the piezo-optomechanical $S_{21}$ taken on a $Y$-cut LN nanobeam with in-plane rotation $\phi=0$ at room temperature. We have chosen this orientation and $\phi$ so that the electrodes only drive the symmetric mechanical bands. The oscillating electric field not only modulates the optical mode frequency via the driven mechanical motion, but also through the electro-optic effect of LN. We repeat the same $S_{21}(\omega)$ measurement with different cavity-laser detuning $\Delta$. The optical cavity response filters the electro-optically generated sidebands, leading to a broad feature with linewidth $\kappa$ in the coherent response. A series of sharper peaks are also visible, and arise from the piezoelectric driving of mechanical resonances of the OMC structure. We also observe a ``clean''  frequency range which matches reasonably well with the simulated mechanical bandgap of the nanobeam mirror cell (shaded region in fig.~\ref{fig4:PiezoDrive}(a)). The localized mechanical modes can be identified near the upper edge of the bandgap. The disordered response outside the bandgap region is not due to noise of the instrument or measurement device, and is observed consistently with the same spectrum. This is likely due to complicated mixing of modes outside of the bandgap.

The mechanical origin of the narrow piezo-optomechanical response peaks can be verified by comparing with the thermal spectroscopy results. Fig.~\ref{fig4:PiezoDrive}(b) shows the piezo-optomechanical response near the upper bandgap edge, including three Fano-shaped peaks. The thermal mechanical noise power spectrum of the same frequency range is shown in fig.~\ref{fig4:PiezoDrive}(c). Fano-shaped response peaks in the piezo-optomechanical $S_{21}$ match closely with the thermal-mechanical noise peaks in the power spectral density. We further calibrate the coherent phonon number from the electrical drive with the thermal phonon occupacy (supplementary material). The decay rate $ \gammae/2\pi = 8.8 \pm 0.56~\textrm{mHz} $ is obtained between the microwave transmission line and the localized mechanical mode. The resulting external microwave-to-mechanics conversion efficiency is $\eta_{\textrm{ext}} = \gammae/\gamma \sim 10^{-8}$. We attribute this low efficiency to the large separation between the coupling electrodes and the impedance mismatch between the electrodes and the $Z_{0}=50~\Omega$ transmission line.

If the OMC were to be connected to a microwave resonator with the same coupling electrode configuration, the microwave-mechanics coupling rate would be $g_{\mu} = \sqrt{\gammae\omegam} \cdot \sqrt{Z_{c}/Z_{0}}/2$, where $Z_{c}$ is the characteristic impedance of the microwave resonator (supplementary material). Based on the measurement of $\gammae$ and assuming $Z_{c} = 50~\Omega$, we estimate $g_{\mu}/2\pi \sim 2~\kilo\textrm{Hz}$. Microwave resonators with impedance on the order of $Z_{c} \sim 5~\kilo\Omega $ can be fabricated with high-kinetic-inductance superconducting nanowires~\cite{samkharadze2016high} and spiral inductors~\cite{barzanjeh2017mechanical}. By optimizing the configuration of the coupling electrodes and engineering the mechanical mode leakage~\cite{arrangoiz2016engineering,patel2017engineering}, strong coupling between the microwave resonator and mechanical mode of the OMC can be realized if $\gammae/2\pi \gtrsim 1~\textrm{Hz}$ and $ Z_{c}/Z_{0} \sim 100 $, giving $g_{\mu}/2\pi \gtrsim 100~\kilo\textrm{Hz} \gtrsim \gamma/2\pi$.

\section{Conclusions}

In summary, we have demonstrated high quality optomechanical crystals on a lithium-niobate-on-silicon platform in the resolved-sideband regime. We measure optical quality factors above $300,000$ and mechanical quality factors as high as $ 17,000 $ at cryogenic temperature. We measure optomechanical backaction, electromagnetic induced transparency and absorption to extract $g_{0} /2\pi\sim 120~\textrm{kHz}$. We observed minor degradation ($\lesssim 15\%$) of the optical linewidth with intracavity photon number $\nc$ up to $1.5\times 10^{5}$. Mechanical lasing is observed with blue-detuned laser pump, corresponding to a cooperativity above unity. Red-detuned measurements with relatively high $\nc$ is not feasible due to the observed quadratic thermal-optic shift, which may be reduced by doping LN with MgO~\cite{furukawa2001green}. Given the strong nonlinear optic effect and electro-optic effect of LN, our high quality LN OMC provides a promising platform for nonlinear photonic application and densely integrated active on-chip photonic and phononic elements.

We further utilize the piezoelectricity of LN to drive the OMC. The mechanical bandgap and localized mechanical modes of the OMC are identified. Based on the measured microwave-to-mechanical conversion rate, we estimate the coupling rate between the OMC and a microwave resonator with identical coupling configuration to be $ g_{\mu}/2\pi \sim 2~\textrm{kHz} $. The microwave-mechanics coupling strength can be improved by reducing the coupling electrodes separation and increasing the characteristic impedance of the microwave resonator. With the recent advance of strong coupling between nanomechanical structures on LN and superconducting circuits~\cite{Arrangoiz-Arriola2018,arrangoiz2019resolving}, a full microwave-to-optical conversion is within reach with our LN OMC platform, opening up opportunities for both classical and quantum signal storing, processing and transduction in hybrid systems exploiting densely integrated electrical, mechanical and optical elements \cite{Safavi-Naeini2019}.

\section*{Funding information}

National Science Foundation (NSF) (1708734, 1808100, 1542152), Army Research Office (ARO/LPS) (CQTS), Office of Naval Research (ONR) (MURI No. N00014-151-2761), Fonds Wetenschappelijk Onderzoek (FWO Marie Sklodowska-Curie grant agreement No. 665501), VOCATIO.

\section*{acknowledgment}

P.A.A. and J.D.W. are partly supported by the Stanford University SGF program. R.N.P. is partly supported by the NSF Graduate Research Fellowships Program. A.S.N. and W.J. thank Martin Fejer and Marek Pechal for numerous helpful discussions. We thank Thaibao Phan supplying us with HSQ. Device fabrication was performed at the Stanford Nano Shared Facilities (SNSF) and the Stanford Nanofabrication Facility (SNF).

\appendix

\part*{Supplementary Information}

\section{Orientation dependency of OMC properties}

\subsection{Definition of coordinate systems and rotated dielectric and photoelastic tensor}

We start by defining the coordinate systems. The global coordinate system is fixed with the nanobeam with axis labeled by $x, y, z$, and the material coordinate system coincides with the crystal axis of LN, denoted by $X, Y, Z$. Note that for Euler angles which rotate the axis of the global coordinate system to the material coordinate system, the corresponding rotation matrix can be applied to transform the tensor components in the material system to the global system.

\begin{figure}[h]
	\includegraphics[scale=0.3]{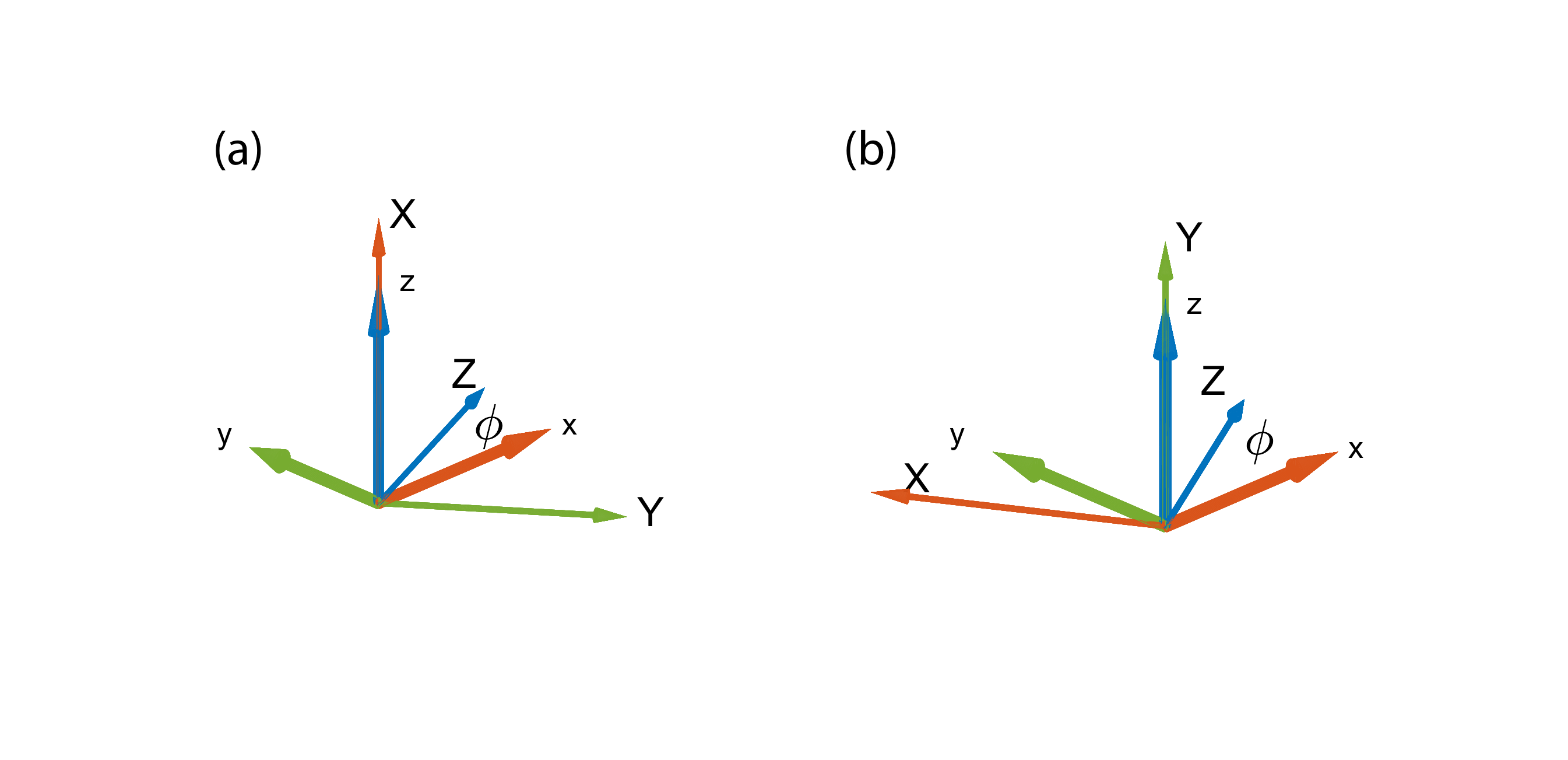}
	\caption{\label{SIFig:CoordSys} Definition of the coordinate systems for (a) X-cut LN and (b) Y-cut LN. The global coordinate system is shown with thicker and shorter arrows, labeled with $x,y,z$. The material coordinate systems are shown with thinner and longer arrows, labeled with $X, Y, Z$. The nanobeam is parallel to the global x axis. The in-plane rotation angle $ \phi $ is defined as the angle between $x$ and $Z$ axis in both case.}
\end{figure}

We first give the rotation matrix used for X-cut LN (LNX) and Y-cut LN (LNY) with in-plane rotation angle $\phi$ as
\begin{eqnarray}
R_{\textrm{LNX}}(\phi) & = & 
\begin{bmatrix}
0 &  \sin \phi  &  \cos \phi  \\
0 & - \cos \phi  &  \sin \phi  \\
1 & 0 & 0
\end{bmatrix},\\
R_{\textrm{LNY}}(\phi) & = & 
\begin{bmatrix}
-\sin \phi  & 0 &  \cos \phi  \\
\cos \phi  & 0 & \sin \phi  \\
0 & 1 & 0
\end{bmatrix}.
\end{eqnarray}
The global coordinate system and the rotated material systems are shown in fig.~\ref{SIFig:CoordSys}. The corresponding Euler angles in `z-x-z' convention are $(\alpha, \beta, \gamma) = (\phi -\pi/2, -\pi/2, -\pi/2) $ for LNX and $(\alpha, \beta, \gamma) = (\phi -\pi/2, -\pi/2, \pi) $ for LNY.

The photoelastic tensor components of the rotated crystal in the global coordinate system are then given by
\begin{equation}
p'_{ijkl}(\phi) = R_{im}(\phi)R_{jn}(\phi)R_{kp}(\phi)R_{lq}(\phi) p_{mnpq}.
\end{equation}
Repeated indices are to be summed. The components of the rotated photoelastic tensor in the global coordinate system are given in Sec.~\ref{SI:RotatedPEComponents}.

For both LNX and LNY, the rotated dielectric tensor is
\begin{equation}
\epsilon' = \begin{bmatrix}
\epsilon _{11} +  \Delta \epsilon_{eo}\cos ^2\phi  & \cos \phi  \sin
\phi \Delta \epsilon_{eo} & 0 \\
\cos \phi  \sin \phi  \Delta \epsilon_{eo} &  \epsilon _{11} +  \Delta \epsilon_{eo}\sin ^2\phi  & 0 \\
0 & 0 & \epsilon _{11}
\end{bmatrix},
\end{equation}
where $ \Delta \epsilon_{eo} =  \epsilon _{33}-\epsilon _{11} $.

\subsection{Simulation of optomechanical coupling rate}

With the rotated dielectric and photoelastic tensor components, the photoelastic contribution of the optomechanical coupling is given by~\cite{chan2012optimized}
\begin{equation}
\label{SIEQ:g0PE}
g_{0,\textrm{PE}} = - \frac{\omegac}{2} \frac{\int \bm{E} \cdot\frac{\partial \bm{\epsilon}}{\partial \alpha} \cdot \bm{E} dV }{\int \bm{E} \cdot \bm{D} dV },
\end{equation}
where $ \alpha$ parametrizes the mechanical motion amplitude, and $ \partial \epsilon_{ij}/ \partial \alpha  = -\epsilon_{ik}\epsilon_{lj}p_{klmn}S_{mn}/\epsilon_{0} $. For isotropic media with refractive index $n$, the photoelastic induced change in dielectric constant simplifies to $ \partial \epsilon_{ij}/ \partial \alpha  = -\epsilon_{0}n^{4}p_{ijmn}S_{mn} $.

Consider the qualitative dependence of photoelastic contribution to the optomechanical coupling rate $g_{0}$. For breathing mechanical modes, the dominating strain component is $S_{yy}$ in the global coordinate system. Similarly, for TE optical modes, the primary electric field component is $E_{y}$. As a result, the largest photoelastic contribution to $g_{0}$ is from $ E_{y}^{2} \partial \epsilon_{yy}/ \partial \alpha \approx -\epsilon_{0}n^{4} p'_{22} S_{yy} E_{y}^{2}  $. This suggests that we focus on the $ p'_{22} $ component of the rotated crystal in the global coordinate system.

\begin{figure}[h]
	\includegraphics[scale=0.6]{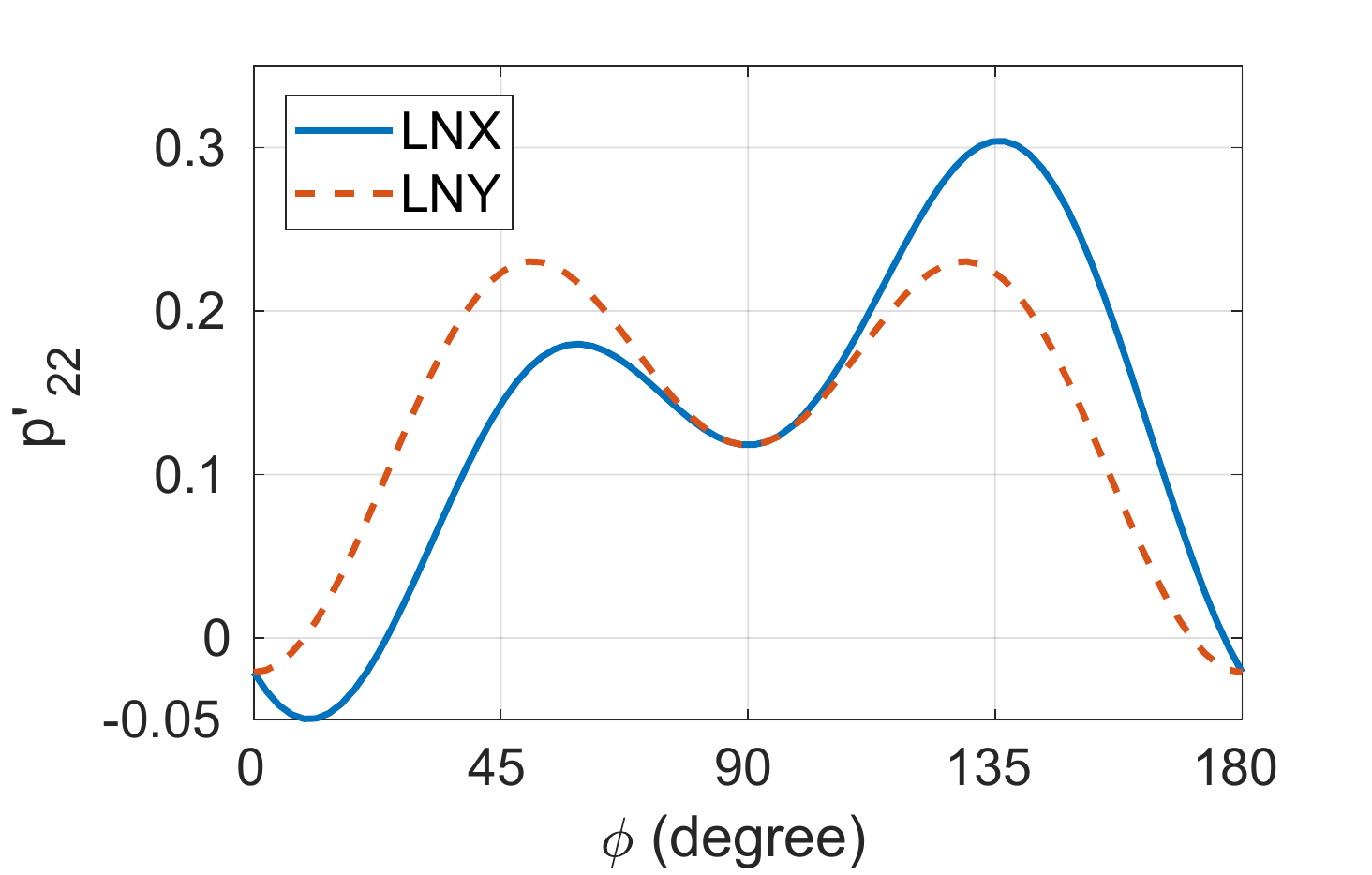}
	\caption{\label{SIFig:p22} Rotated photoelastic tensor component $ p'_{22}$ for LNX (solid blue) and LNY (dashed red). }
\end{figure}

We use photoelastic components from Ref.~\cite{andrushchak2009complete} for numerical evaluations. The $p'_{22}$ component of LNX and LNY is plotted in fig.~\ref{SIFig:p22}. The maximal $p'_{22}$ can be achieved on LNX with $ \phi = 135$ degree, where $p'_{22} = (p_{11}+p_{33} +p_{13}+p_{31})/4+p_{44}-(p_{14}+p_{41} )/2 \approx 0.3$.

For the moving boundary contribution of the optomechanical interaction, we approximate LN as an isotropic dielectric material with refractive index $n = \sqrt{\epsilon_{\textrm{LN}}}= 2.2$. The contribution from moving boundary is~\cite{chan2012optimized}
\begin{equation}
\label{SIEQ:g0MB}
g_{0, \textrm{MB}} = -\frac{\omegac}{2} \frac{\oint (\bm{Q}\cdot \bm{n}) (\Delta \epsilon \bm{E}^{2}_{\parallel} -\Delta \epsilon^{-1} \bm{D}^{2}_{\perp})dS }{\int \bm{E} \cdot \bm{D} dV },
\end{equation}
where $\Delta \epsilon \equiv \epsilon_{\textrm{LN}} - \epsilon_{\textrm{air}}  $, $\Delta \epsilon^{-1} \equiv \epsilon_{\textrm{LN}}^{-1} - \epsilon_{\textrm{air}}^{-1}  $, $\bm{Q}$ is the normalized displacement field and $\bm{n}$ is the surface norm pointing towards the air. The subscripts $\parallel$ and $\perp$ denote the parallel and perpendicular component of the fields locally with respect to the surface.

\begin{figure}[h]
	\includegraphics[scale=0.35]{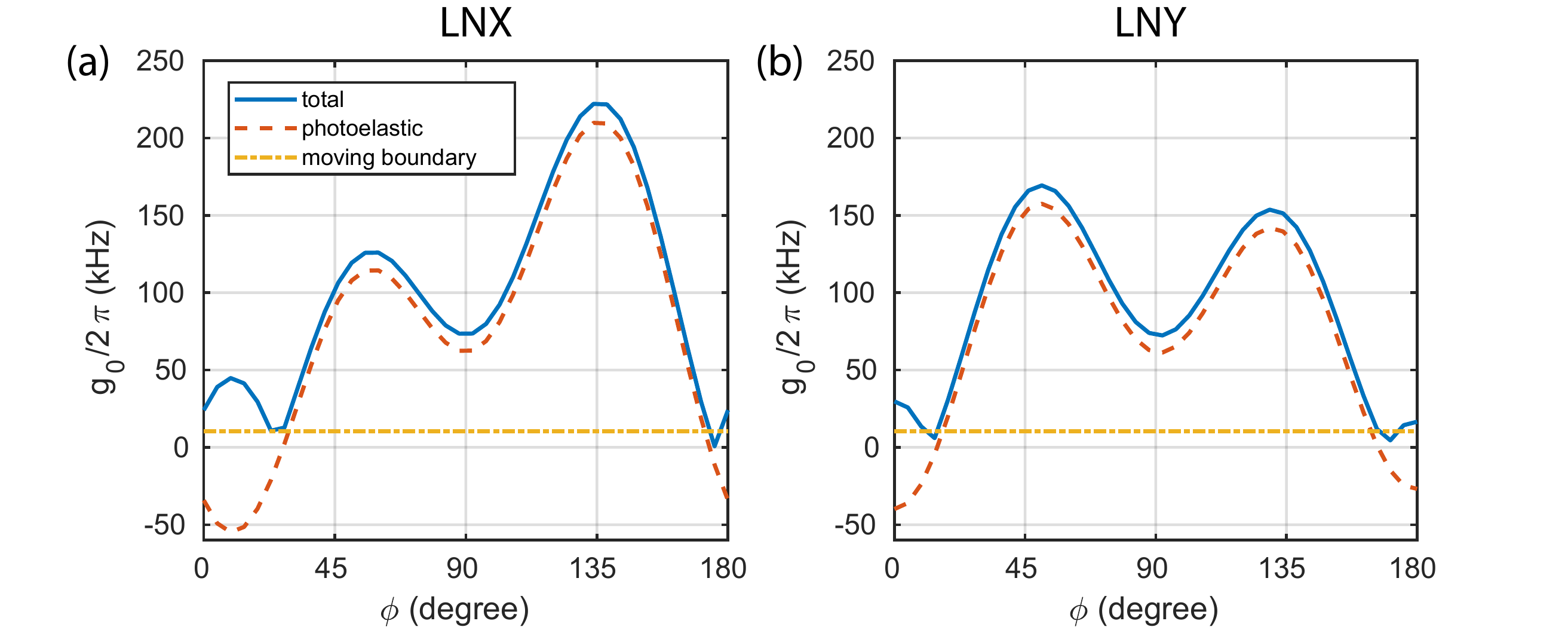}
	\caption{\label{SIFig:g0sim} Simulation of optomechanical coupling rate $g_{0}$ on (a) $X$-cut and (b) $Y$-cut LN for different in-plane rotation angle $\phi$. }
\end{figure}

We assume same mode profiles for different $\phi$ as a first-order approximation to calculate the $\phi$-dependence of the optomechanical coupling rate $g_{0}$. We evaluate eq.~\ref{SIEQ:g0PE} and eq.~\ref{SIEQ:g0MB} for different $\phi$. In fig.~\ref{SIFig:g0sim} we show the photoelastic contribution $g_{0, \textrm{PE}}$ (dashed red), moving boundary contribution $g_{0, \textrm{MB}}$ (dashed dotted yellow) and the absolute values of the total $g_{0}$ (blue). From fig.~\ref{SIFig:p22}, it is clear that $p'_{22}$ is a good indicator of the optomechanical coupling rate.

\subsection{Measurements of optical quality factors and optomechanical coupling rates with various $\phi$}

\begin{figure*}[h]
	\includegraphics[scale=0.58]{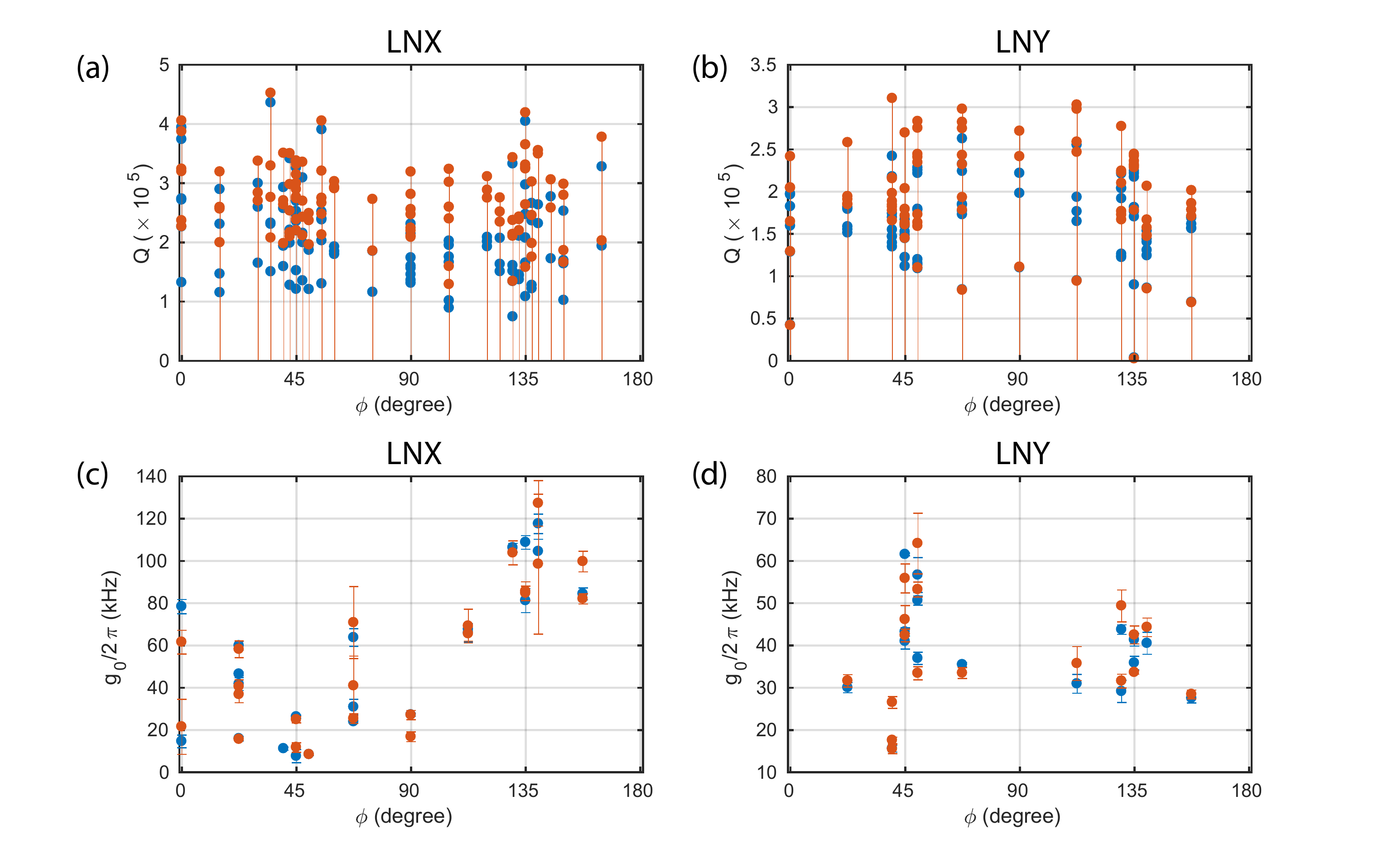}
	\caption{\label{SIFig:Q_and_g} Optical quality factor $Q$ and optomechanical coupling rate $g_{0}$ on LNX and LNY. (a, b) Measured total (blue) and intrinsic (red) optical quality factors on LNX (a) and LNY (b). No obvious dependency on in-plane rotation $\phi$ can be observed. The quality factors on LNY are generally lower due to fabrication variations between different chips. (c, d) Measured zero-point optomechanical coupling rate $g_{0}$ for different $\phi$ on LNX (c) and LNY (d). Blue (red) data points represent measurement with detuning $\Delta = -\omegam$ ($\Delta = \omegam $). Error bars represent one standard deviation obtained by repeated measurements with different laser pump power.}
\end{figure*}

We fabricated OMCs on both LNX and LNY with various values of $\phi$. The measured optical quality factors are shown in fig.~\ref{SIFig:Q_and_g}(a, b). No obvious dependence of quality factors on $\phi$ is observed. The quality factors on LNY are generally lower due to fabrication variations between different chips. The mechanical quality factors at room temperature are typically $Q_{\textrm{m}}\sim 4000$. The room-temperature $Q_{\textrm{m}}$ is limited by thermoelastic damping and is relatively insensitive to crystal orientations. At cryogenic temperature, the highest mechanical $Q_{\textrm{m}} = 37,000$ is observed on the LNY OMC with $\phi = 0$, where the y-symmetry of the nanobeam is recovered.

In fig.~\ref{SIFig:Q_and_g}(c, d) we show the measured optomechanical coupling rates $g_{0}$ versus OMC orientation angle $\phi$. The coupling rates were measured at room temperature ($300~\textrm{K}$) using calibrated thermal mechanical noise power spectral density~\cite{patel2018single}. The measured $g_{0}$'s are in general roughly smaller than the simulated values by $\sim 50\%$. We attribute this discrepancy to the approximation of mode profiles, the fact that the design is only optimized for $\phi=0$ on LNY, the possible differences in the material properties between simulation and the actual LNOS wafers, and the considerable uncertainties of LN's photoelastic components~\cite{andrushchak2009complete}.

Despite the discrepancy between the absolute values of the simulated and measured $g_{0}$,  we observe that the maximal optomechanical coupling rate occurs at $ \phi = 135^\circ$ on LNX, agreeing with the simulations (fig.~\ref{SIFig:g0sim}(a)) and also the simple prediction from $p'_{22}$ (fig.~\ref{SIFig:p22}). The optimal orientations on LNY are near $\phi = 45^\circ$ and $\phi=135^\circ$, reasonably matching the simulation results in fig.~\ref{SIFig:g0sim}(b).

\section{Thermal-optical shift and thermal relaxation}
\label{sec:thermal-optical}

\subsection{Thermal-optical shift}
\label{subs:thermal-optical-shift}

\begin{figure*}[h]
	\centering
	\includegraphics[scale=0.6]{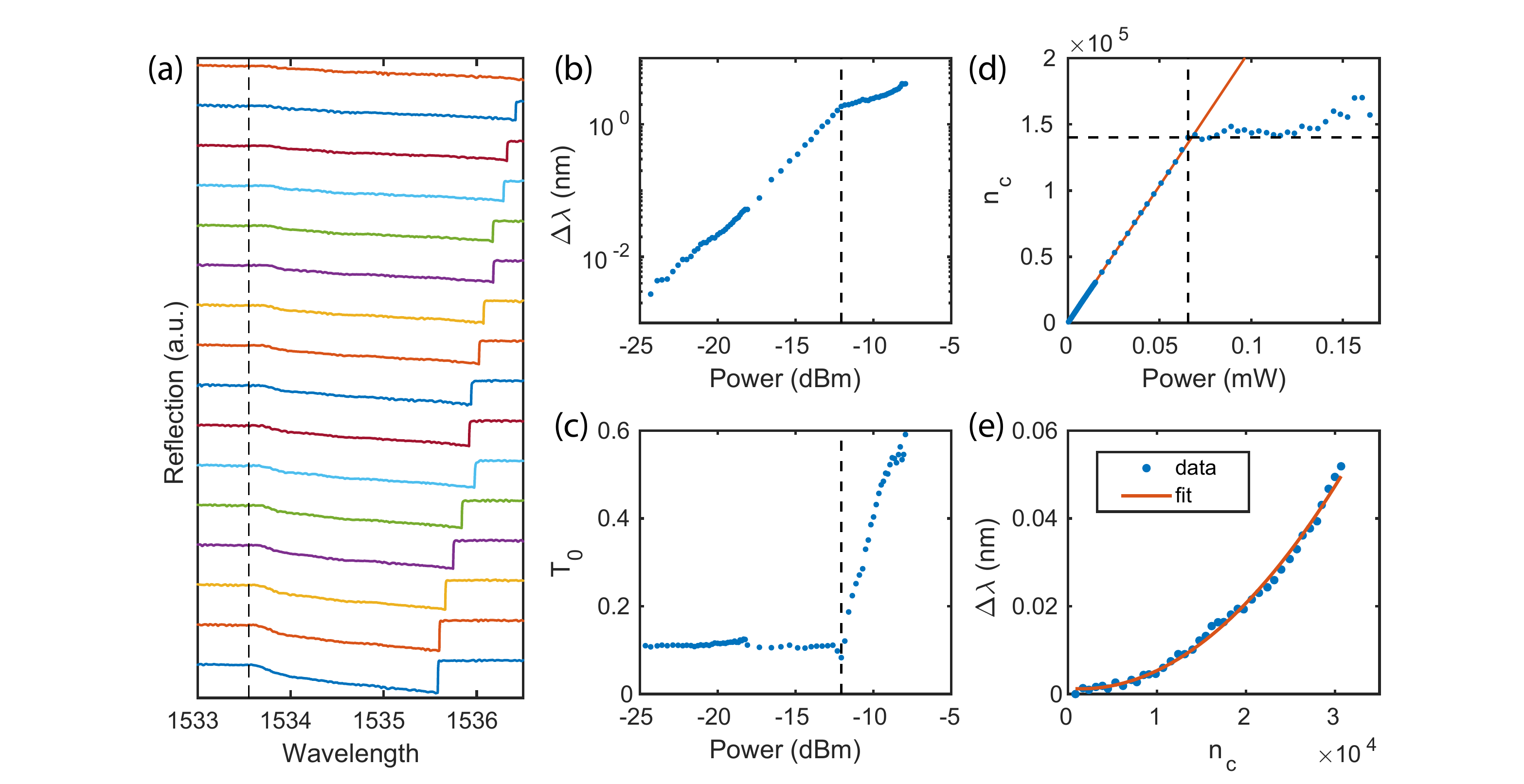}
	\caption{\label{SIFig:ThermShift} Thermal-optical shift measurements. (a) Reflection signal of laser wavelength scan at increasing powers. Vertically displaced for viewing purposes. The vertical dashed line represents the cavity wavelength at low power. (b) Maximum optical cavity wavelength shift versus laser power, extracted from (a). (c) Minimum reflection $T_{0}$ at the resonance versus laser power, extracted from (a). (d) Maximal intracavity photon number versus laser power. (e) Wavelength shifts versus $\nc$ before saturation (blue dots) and a quadratic fit (red line). }
\end{figure*}

We observed a thermal-induced optical shift when scanning the laser over the optical cavity with different powers. The thermal-induced optical shift is well understood in silicon microcavities~\cite{carmon2004dynamical}. To better understand the thermal-optical shift on LN OMC, we briefly describe the silicon case here.

In silicon microcavities, the heat absorption rate is proportional to the intracavity photon number $\nc$, leading to a local temperature change proportional to $\nc$. The temperature change affects the optical mode via thermal expansion and temperature-dependent refractive index change, both result in a wavelength shift that's linear with respect to the temperature change. A positive wavelength shift is usually observed for an increase in temperature. When the laser frequency approaches the optical cavity from the blue side, $\nc$, the temperature increase, and the optical cavity shifts red. When the laser reaches beyond the maximum wavelength shift which occurs for maximum $\nc$ at the reflection dip, $\nc$ starts decreasing. The cavity shifts back towards the blue side as a result of decreasing $\nc$, which further decreases $\nc$. This causes the cavity to jump back to nearly its original wavelength. The cavity red shift reflects itself in the continuous blue-to-red laser wavelength sweep as a slow slope which gets flatter for higher power. The mode escapes after the laser has passed the maximum shifted wavelength, giving a sharp jump in the transmission or reflection spectrum similar to Fig~\ref{SIFig:ThermShift}(a).

The thermal-induced optical shift $\Delta \lambda$ we observed on the LN OMC is qualitatively similar to silicon, but the wavelength shifts faster than linear and could be as large as few nanometers at high laser powers. As shown in fig.~\ref{SIFig:ThermShift}(a), the reflections were measured for linearly increasing laser scan powers, vertically displaced for viewing purposes. We extract the maximum wavelength shifts $\Delta \lambda_{\textrm{max}}$ (fig.~\ref{SIFig:ThermShift}(b)) as well as minimum reflections at the resonance dip $T_{0}$ versus laser powers (fig.~\ref{SIFig:ThermShift}(c)). A sharp transition occurs near a threshold laser power $P_{\textrm{in, thres}} = 0.065~\textrm{mW}$. The wavelength shifts slow down and the transition dips get shallower.

With the coherent spectroscopy method, we confirmed that with powers $P_{\textrm{in}} > P_{\textrm{in, thres}} $ and $\nc$ as high as $1.5\times 10^{5}$, the cavity linewidth stays roughly constant regardless of the increasing $T_{0}$ (see sec.~\ref{sec:kappa-vs-nc}). Little additional loss is introduced by the high optical powers used in our measurements. We conclude that the increase in $T_{0}$ is due to an ``early escape'' of the mode, where $\Delta \lambda$ saturates and the cavity escapes back to far blue before the laser reaches the reflection dip. Based on measurements of the reflection dip $T_{0} = |(i\Delta + \kappa/2 - \kappae)/(i\Delta + \kappa/2)|^{2}$, we could obtain the detuning $\Delta$ right before the mode escapes and calculate the corresponding intracavity photon number $\nc$. This $\nc$ is the maximal intracavity photon number obtained for each laser power. The results are shown in fig.~\ref{SIFig:ThermShift}(d). For laser powers below $P_{\textrm{in, thres}}$, the maximal $\nc$ grows linearly with respect to input powers. When $\nc$ reaches $n_{c, \textrm{thres}}\sim 1.4\times 10^{5}$ at $P_{\textrm{in}} = P_{\textrm{in,thres}}$, the intracavity photon numbers are observed to saturate at values near $n_{c,\textrm{thres}}$, largely deviating from the linear relation (red line) regardless of increasing laser powers. Similar threshold behaviors in terms of $\nc$ were observed on different OMC devices and showed no obvious correlation with the optical quality factors. The physical mechanism leading to the threshold and the dependence of $n_{c, \textrm{thres}} $ on system parameters are not understood and will be the subject of further exploration.

For $\nc$ smaller than the threshold, we plot the wavelength shifts versus the intracavity photon numbers before $n_{c,\textrm{thres}}$ in fig.~\ref{SIFig:ThermShift}(e). A quadratic relationship between wavelength shift and intracavity photon number is obtained with $\Delta \lambda = \alpha_{2} \nc^{2}$ where we call $ \alpha_{2}$ the quadratic thermal-induced cavity shift coefficient. $\alpha_{2} \sim 6\times 10^{-11}~\nano\meter/(\textrm{photon})^2 $ is obtained from the fit. The cavity shift could also be expressed in terms of frequency shift $\Delta \omegac = \alpha_{2} \nc^{2} $ where $\alpha_{2}\sim  -7~\textrm{Hz}/(\textrm{photon})^2$. We observed that the measured $\Delta \lambda$ is well characterized by the quadratic fit for $10^{4} \lesssim \nc \lesssim 10^{5}$ and deviates from the fit at low and high power, showing that the actual thermal-induced cavity shift has a complicated dependency on $\nc$, including a small linear $\nc$ term and also terms with higher order. The quadratic contribution already dominates the wavelength shift when the shift is noticeable with $\Delta \lambda \gtrsim \kappa$, making the linear term difficult to measure.

\subsection{Measurement of the thermal relaxation rate}

In this section we consider the response of the cavity to a laser pump with slow and weak amplitude modulation. We first define related parameters and variables in Table~\ref{tab:parameters}. When referring to the steady-state values at a constant power, quantities are denoted with a bar.

\begin{table*}
	\caption{Definition of parameters related to thermal relaxation rate measurements.} \label{tab:parameters}
	\begin{center}
		\begin{tabular}{ccc}
			Parameter & Description & Useful relation \\ 
			\hline
			$\Pin$ & input laser power& \\
			$\Delta T$ & temperature change &  \\
			$\Delta \omegac$ & thermal-induced cavity shift& $\Delta \omegac = \alpha_{1} \nc + \alpha_{2} \nc^{2} $ \\ 
			$\Delta$ & real-time cavity-laser detuning & $\Delta = \omegac - \omega_{\textrm{l}} + \Delta \omegac$ \\
			$\dot{N}_{\textrm{in}}$ & input photon flux & $ \Pin/(\hbar \omega_{\textrm{l}}) $\\
			$\nc $ & intracavity photon number &  $\nc = \kappae \dot{N}_{\textrm{in}} /(\Delta ^2 + (\kappa/2)^2) $\\
			$\delta \Pin$ & small laser power variation & \\
			$ \delta \nc $ & small $\nc $ variation due to $\delta \Pin$ & \\
			$\delta \omegac $ & extra cavity shift variation due to  $\delta \nc$ & \\
			$\delta \Delta$ & small detuning variation &  $\delta \Delta = \delta \omegac $ for fixed laser wavelength\\
			$\Gamma$ & temperature relaxation rate & \\
			$\Gamma_{2}$ & measurement induced extra relaxation rate & \\
			$\Gammatot$ & total measured relaxation rate & $\Gammatot = \Gamma + \Gamma_{2}$
		\end{tabular}
	\end{center}
\end{table*}

We start by assuming a linear thermal-optical shift of the optical cavity frequency with respect to temperature change
\begin{equation}
\Delta \omegac = C_{0} \Delta T.
\end{equation}

Then dynamics of the temperature change is modeled by
\begin{equation}
\frac{d \Delta T }{d t} = - \Gamma \Delta T + f(\nc),
\end{equation}
where $\nc = \kappae \dot{N}_{\textrm{in}}/(\Delta^{2} + (\kappa /2)^{2})$ is the instant intracavity photon number and $\dot{N}_{\textrm{in}} = P_{\textrm{in}}/(\hbar \omegac)$ is the input photon flux of the pump. An exponential temperature relaxation with rate $\Gamma$ is assumed. We also assume that the heating effect only directly depends on intracavity photon number $\nc$ via a general function $f$. The dynamical equation could  be formulated in terms of the cavity shift $\Delta \omegac$ as
\begin{equation}
\label{EQSI:DeltaDynamics}
\frac{d \Delta \omegac  }{d t} = - \Gamma \Delta \omegac + C_{0} f(\nc).
\end{equation}
This equation describes the complicated dynamics of $\Delta \omegac $ which also enters $f(\nc)$ implicitly via $\nc$. $C_{0}$ can be absorbed into the function $f$ so that $f$ has the dimension of $(\textrm{rad/s})^2$. We will omit $C_{0}$ from now on. Note that eq.~\ref{EQSI:DeltaDynamics} gives the steady-state cavity shift $\overline{\Delta \omegac} = f(\ncsty)/\Gamma$ from which we could solve $\overline{\Delta \omegac} $ and $\ncsty$ if $f$ and $\Gamma $ are known.

Based on eq.~\ref{EQSI:DeltaDynamics}, we consider small time-dependent variations of laser power $\delta P_{\textrm{in}}$ near the steady state solution. The laser wavelength is kept fixed. Note that $\nc$ is a function of $\Delta$ and $P_{\textrm{in}}$ so that
\begin{eqnarray}
\frac{d \delta \Delta}{d t}  &=& -\Gamma (\overline{\Delta \omegac} + \delta \Delta) + f(\ncsty)+ f'(\ncsty)\cdot \delta \nc\\
& = & -\Gamma \delta \Delta + f'(\ncsty)\cdot \left(\frac{ \ncsty \delta \Pin}{\overline{\Pin}} - \frac{2 \ncsty \bar{\Delta} \delta \Delta}{\bar{\Delta}^{2} + (\kappa /2)^{2}}  \right)\\
\label{EQSI:deltaDeltaDynamics}& = & - (\Gamma + \Gamma_{2}) \delta \Delta + f'(\ncsty)\ncsty \cdot \frac{\delta \Pin}{\overline{\Pin}},
\end{eqnarray}
where
\begin{equation}
\label{EQSI:Gamma2}
\Gamma_{2} = \frac{2  \bar{\Delta}}{\bar{\Delta}^{2} + (\kappa /2)^{2}} \cdot f'(\ncsty)  \ncsty = g(\bar{\Delta}) \cdot h(\ncsty).
\end{equation}
The extra relaxation rate $\Gamma_{2}$ reflects the fact that the laser amplitude-modulation measurement method enters the complicated dynamics of the system itself, leading to an effective relaxation rate in addition to the actual relaxation rate $\Gamma$. We define functions $g$ and $h$ to represent the dependency of $\Gamma_{2} $ on $\bar{\Delta}$ and $\ncsty$. We point out that $\bar{\Delta}$ and $\ncsty$ can be independently controlled by preparing the steady state with different laser frequencies $\omega_{\textrm{l}}$ and powers $\Pin$.

When the input laser power has a small variation $\delta \Pin = \alpha \cos (\omega t)$, by assuming $\delta \Delta = \beta \cos (\omega t + \phi)$ and plugging into eq.~\ref{EQSI:deltaDeltaDynamics}, we get the amplitude response of $\delta \Delta$ to $\delta \Pin$ to be
\begin{equation}
\left | \frac{\beta}{\alpha} \right| \propto \frac{1}{\sqrt{\omega^{2} + (\Gamma + \Gamma_{2})^2}}.
\end{equation}
As a result, the total ``relaxation rate'' $\Gammatot = \Gamma + \Gamma_{2}$ can be measured from the low-frequency amplitude response of $\delta \Delta$. With the measurement of $\Gammatot$ and independent control over $\bar{\Delta}$ and $\ncsty$, we are allowed to probe the structure of $\Gamma_{2}$ and thus the unknown function $f$.

Proceeding to the determination of $f$, we assume a polynomial form $f(n) = a_{1}n + a_{2}n^{2} + O(n^{3})$, where higher order terms are assumed to be small based on the steady-state cavity shift measurements. With this assumption, $h(n) = nf'(n)= a_{1}n + 2 a_{2}n^{2}$, and
\begin{equation}
\label{EQSI:LinReg}
\Gammatot = (1,g(\bar{\Delta})\ncsty , 2g(\bar{\Delta}) \ncsty^{2} )\cdot (\Gamma, a_{1}, a_{2})^{T} = \bm{M}\cdot \bm{b}.
\end{equation}
By carrying out multiple measurements of $\Gammatot$ with different $\bm{M} $, the thermal relaxation rate $\Gamma$ and coefficients $a_{1}$ and $a_{2}$ can be determined by solving a linear regression problem. Once $\Gamma, a_{1}$ and $a_{2}$ are obtained, the steady-state cavity shift is $ \overline{\Delta \omegac} = \alpha_{1} \ncsty + \alpha_{2} \ncsty^{2} = a_{1}\ncsty/\Gamma + a_{2}\ncsty^{2}/\Gamma $.

We now show that the response of $\delta\Delta$ can be directly measured by the same coherent spectroscopy setup at low modulation frequencies. When there is no slow thermal-induced cavity wavelength shift, the optical response is given by
\begin{equation}
r(\omega \ll \Delta) \approx 1 - \frac{\kappae}{i(\Delta - \omega) + \kappa/2}.
\end{equation}
When the slow cavity wavelength shift is considered, $\Delta$ is replaced by $\bar{\Delta} + \delta\Delta(\omega)$, where $\delta \Delta(\omega)$ is the frequency-domain response of the slow thermal-induced wavelength shift. For small and low-frequency intensity modulation, we assume $\omega \ll \delta\Delta \ll \Delta$, and the response can be simplified as
\begin{eqnarray}
r(\omega \ll \Delta) &\approx& 1 - \frac{\kappae}{i(\Delta+\delta\Delta (\omega)) + \kappa/2} \\
&\approx& 1 - \frac{\kappae}{i\Delta + \kappa/2} + \frac{\kappae}{i\Delta + \kappa/2} \cdot \frac{i\delta\Delta(\omega)}{i\Delta + \kappa/2}.
\end{eqnarray}
The resulting intensity response is
\begin{equation}
|r(\omega \ll \Delta)|^2 =A + B\cdot \delta \Delta(\omega) + O \left((\frac{\delta\Delta}{\Delta})^2 \right),
\end{equation}
where $A$ and $B$ are two real-valued functions of $\Delta$, $\kappa$ and $\kappae$. The exact forms of $A$ and $B$ are involving but not important since they are approximately independent of frequency for $\omega \ll \Delta $. As a result, for low frequency intensity modulation, the only frequency-dependent intensity response is from the variation of thermal-induced cavity shift $\delta \Delta = \delta \omegac$.

\begin{figure}[h]
	\includegraphics[scale=0.27]{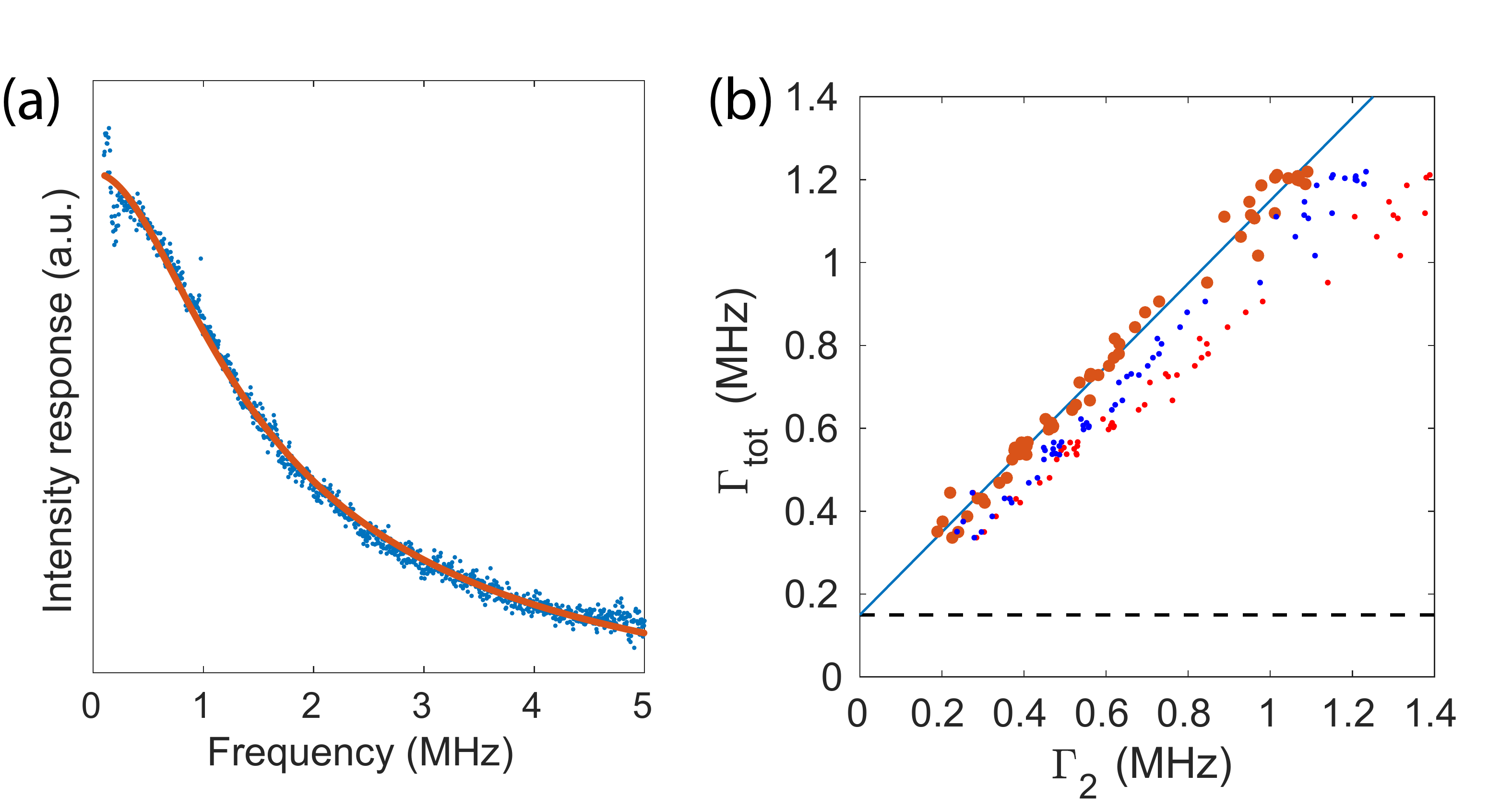}
	\caption{\label{SIFig:ThermRelax} Measurement of thermal relaxation rate. (a) Example of low frequency intensity modulation response of the OMC system. Blue: data. Red: Lorentzian fit. (b) Measured total relaxation rate versus measurement induced extra relaxation rate (orange). The black dashed line shows the actual thermal relaxation rate of the OMC. The blue line is a guide for the eye and has a slope of 1. }
\end{figure}

We measured the intensity response $|r(\omega\ll \Delta)|^2$ for frequency range $100~\kilo\textrm{Hz} \sim 5~\mega\textrm{Hz}$. The amplitude of the response is fitted to a Lorentzian centered at frequency $f=0$. The linewidth of this Lorentzian corresponds to the total relaxation rate $\Gammatot$. A typical low frequency response is shown in fig.~\ref{SIFig:ThermRelax}(a) with the Lorentzian fit (red). In this way we extracted $\Gammatot$ under different laser powers and wavelengths. The steady-state cavity-laser detuning $\bar{\Delta}$ is measured at the same time by an optical sideband sweep across the cavity. With measured $\Gammatot$, $\bar{\Delta}$ and $\bar{\nc}$, the linear regression problem eq.~\ref{EQSI:LinReg} is solved. We obtain the true thermal relaxation rate $\Gamma = 150~\textrm{kHz}$, linear thermal-induced cavity shift coefficient $\alpha_{1} = a_{1}/\Gamma \approx -7.1\times 10^{4}~\textrm{Hz}/\textrm{photon}$ and quadratic thermal-induced cavity shift coefficient $\alpha_{2} = a_{2}/\Gamma \approx -1.5~\textrm{Hz}/(\textrm{photon})^{2}$. 

The thermal relaxation rate $\Gamma$ quantitatively agrees with the thermal-optical response time scale of $\sim 10~\micro s$ reported in Ref.~\cite{sun2017nonlinear} and is slightly faster due to smaller device volume. We note that for $\nc \gtrsim 5\times 10^{4}$, the thermal-optical shift starts to be dominated by the quadratic contribution, which agrees with the steady-state wavelength shift measurement in the last section. We note that the $\alpha_{2}$ obtained here is smaller than the value from steady-state wavelength shift measurement by a factor $ \sim 4$.

We further calculated $\Gamma_{2}$  with $a_{1}$ and $a_{2}$ from the linear regression and $\nc$ from the measurements. We show in fig.~\ref{SIFig:ThermRelax}(b) the measured $\Gammatot$ versus calculated $\Gamma_{2}$ (orange dots). The horizontal black dashed line corresponds to the constant $\Gamma$. A blue line starting at $(0, \Gamma)$ with slope equals to one is plotted for guide of the eye. It's clear that good agreement is obtained between the linear regression results and the equation $\Gammatot = \Gamma + \Gamma_{2}$, under a large variation of both laser powers  and cavity-laser detunings. To show the reliability of the linear regression results, we manually increase the obtained value of $a_{1}$ ($a_{2}$) by $50\%$ and plot the modified $\Gamma_{2}$ in fig.~\ref{SIFig:ThermRelax}(b) as small blue (red) dots for comparison. Deviation from $\Gammatot = \Gamma + \Gamma_{2} $ can be clearly observed.

\section{Measurement of optical linewidth with different intracavity photon numbers}
\label{sec:kappa-vs-nc}

In sec.~\ref{sec:thermal-optical}\ref{subs:thermal-optical-shift} we observed the minimal transmission $T_{0}$ increased for high laser powers. We further deduced the minimal cavity-laser detunings before the optical mode jumped back to the blue side of the laser based on the minimal transmission measurements. However, a change in the cavity intrinsic linewidth $\kappai$ could also change the transmission dip, where $ T_{0}|_{\Delta=0} = |(\kappa/2 - \kappae)/(\kappa/2)|^{2} =|(\kappai - \kappae)/(\kappai + \kappae)|^{2} $. To track any considerable change in the cavity intrinsic linewidth, we fit the probe response from the coherent spectroscopy with different pump laser powers and different cavity-laser detunings. This two-tone spectroscopy effectively gives us the linear response of the cavity at different pump powers and helps us determine $\Delta$, $\kappae$, and $\kappai$, as a function of laser power and detuning.

\begin{figure}[h]
	\includegraphics[scale=0.37]{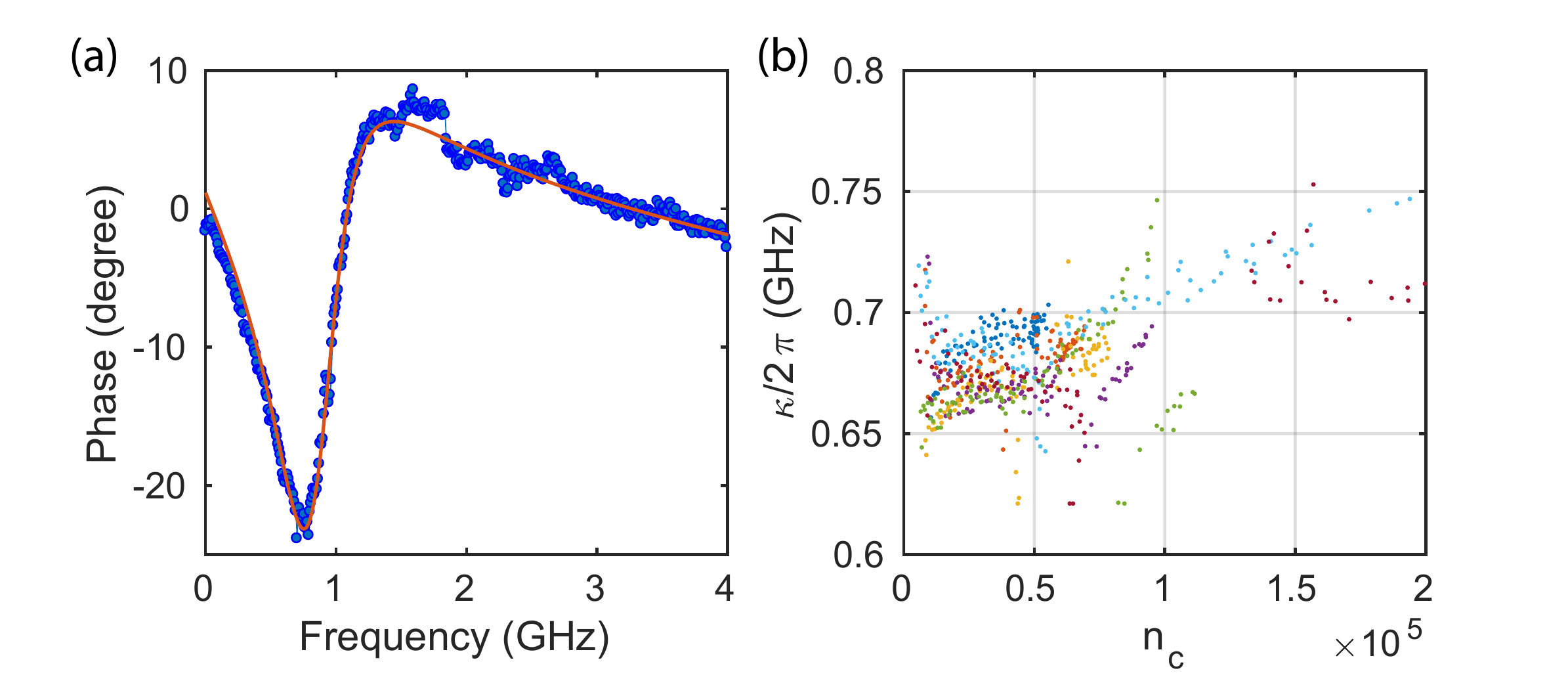}
	\caption{\label{SIFig:kappa-vs-nc} Measurement of optical linewidth $\kappa$ with different intracavity photon numbers $\nc$. (a) A typical phase response of the probe with a high pump power $P_{\textrm{in}} \sim 0.2~\textrm{mW}$ and a near-cavity detuning $\Delta \sim \kappa$. The red line shows the fit result. (b) Extracted total optical linewidth $\kappa$ versus different intracavity photon numbers $\nc$ under different pump powers and detunings. Different colors correspond to different pump laser powers.}
\end{figure}

Fig.~\ref{SIFig:kappa-vs-nc}(a) shows one typical phase response of the probe with a high pump laser power $P_{\textrm{in}} = 0.21~\textrm{mW}$ and a small cavity-laser detuning $\Delta \sim -\kappa$. By fitting the phase response (red), we extract the total optical linewidth $\kappa$. Measurements of $\kappa$ versus intracavity photon numbers $\nc$ with various pump powers and detunings are plotted in fig.~\ref{SIFig:kappa-vs-nc}(b). Different colors correspond to different pump laser powers ranging from $96~\micro\textrm{W}$ to $0.21~\textrm{mW}$. We obtained $\kappa \approx 0.7~\textrm{GHz} $ with relative standard deviation $\sim 9\%$ for all measurements with different pump powers. A linear increase in total linewidth for increasing $\nc$ can be observed and shows different slopes for different pump powers. We note that in general the optical linewidth varies by less than $ 15\%$ for $\nc$ varying more than two orders of magnitude.

\section{Coupling and conversion between microwave, mechanics and optics}

\subsection{Input-output formalism of a optomechanical crystal coupling to a microwave resonator or a microwave  channel}

Consider a system with an OMC coupling to a microwave resonator $c$ with coupling strength $\gmu$. We start by writing down the frequency-domain Heisenberg-Langevin equations of motion in a rotating frame at the laser frequency for $ \Delta \sim \omegam $:
\begin{eqnarray}
- i\omega a(\omega) &=& -(i \Delta + \frac{\kappa}{2}) a - i G b - \sqrt{\kappae} \ain \\
- i \omega b(\omega) &=& -(i \omegam + \frac{\gamma}{2}) b - i G a  - i \gmu c- \sqrt{\gammae} \bin \\
- i \omega c(\omega) &=& -(i \omega_{\mu} + \frac{\kappamu}{2}) c - i \gmu b - \sqrt{\kappamue} \cin
\end{eqnarray}
We assume that every mode is coupled to a single external channel respectively. The counter-rotating terms have been omitted for simplicity. We introduce short-hands
\begin{eqnarray}
\Aa &=& i(\Delta - \omega) + \kappa/2\\
\Ab &=& i(\omegam - \omega) + \gamma/2\\
\Ac &=& i(\omega_{\mu} - \omega) + \kappamu/2\\
\etaab &=& -iG/\Aa \\
\etaba &=& -iG/\Ab \\
\etabc &=& -ig/\Ab \\
\etacb &=& -ig/\Ac \\
\eta_{ijk} &=& \eta_{ij} \eta_{jk}
\end{eqnarray}
where $i,j,k$ run through $a,b,c$. With them the solution can be expressed as
\begin{equation}
\label{eqnSI:Solved_abc}
\begin{bmatrix}
a\\
b\\
c
\end{bmatrix} = \frac{1}{1-\etaaba-\etabcb } \begin{bmatrix}
1- \etabcb & \etaab & \etaabc\\
\etaba & 1 & \etabc\\
\etacba & \etacb & 1 -\etaaba
\end{bmatrix} 
\begin{bmatrix}
\frac{-\sqrt{\kappa_e} \ain}{\Aa}\\
\frac{-\sqrt{\gamma_e} \bin}{\Ab}\\
\frac{-\sqrt{\kappamue} \cin}{\Ac}
\end{bmatrix}
\end{equation}

From eq.~\ref{eqnSI:Solved_abc}, we can directly read the electro-optic conversion $S$ parameters when only input $\cin$ or $\ain$ presents:
\begin{eqnarray}
S_{\textrm{ac}} &\equiv& \frac{\aout}{\cin}  = \frac{\sqrt{\kappae} a}{\cin}  \\
& = & \sqrt{\kappae}\frac{\etaabc}{1-\etaaba -\etabcb} \frac{-\sqrt{\kappamue}}{\Ac}\\
&=& \sqrt{\kappae} \frac{1}{\Aa} \frac{G \gmu}{ \Ab + G^{2}/\Aa + \gmu^{2}/\Ac }\frac{1}{\Ac} \sqrt{\kappamue}\\
&=& S_{\textrm{ca}} \equiv \frac{\cout}{\ain}
\end{eqnarray}
At perfectly matched frequencies $\omega= \Delta = \omegam = \omega_{\mu} $, we have the conversion efficiency
\begin{eqnarray}
\eta &\equiv& \left| S_{\textrm{ac}} \right|^{2} =  \left| S_{\textrm{ca}} \right|^{2}\\
&=& \kappae \frac{4}{\kappa^{2}} \frac{ 4G^{2} \gmu^{2}/\gamma^{2} }{ (1+ 4G^{2}/\kappa\gamma + 4\gmu^{2}/\kappamu\gamma )^{2} } \frac{4}{\kappamu^{2}} \kappamue\\
&=& \eta_{\textrm{ext,a}}\eta_{\textrm{ext,c}} \frac{4\Cab \Cbc}{(1+\Cab + \Cbc)^{2}}
\end{eqnarray}
where $\eta_{\textrm{ext,a}} = \kappae/\kappa $ and $ \eta_{\textrm{ext,c}} =\kappamue/\kappamu$ are defined as the external efficiencies. $\Cab$ ($\Cbc$) is the cooperativity between the mechanical mode and the optical (microwave) mode. The maximal conversion efficiency $ \eta = \eta_{\textrm{ext,a}}\eta_{\textrm{ext,c}} $ is achieved for $ \Cab = \Cbc \gg 1 $.

We proceed to consider the simplified case where the mechanical mode is directly coupled to a microwave channel. The external decay rate $\gammae$ now represents the coupling between the microwave channel and the mechanical mode. The equations of motion read
\begin{eqnarray}
- i\omega a(\omega) &=& -(i \Delta + \frac{\kappa}{2}) a - i G b - \sqrt{\kappae} \ain \\
- i \omega b(\omega) &=& -(i \omegam + \frac{\gamma}{2}) b - i G a - \sqrt{\gammae} \cin
\end{eqnarray}
and the simplified version of eq.~\ref{eqnSI:Solved_abc} is
\begin{equation}
\label{eqnSI:Solved_ab}
\begin{bmatrix}
a\\
b
\end{bmatrix} = \frac{1}{1-\etaaba } \begin{bmatrix}
1 & \etaab \\
\etaba & 1 
\end{bmatrix} 
\begin{bmatrix}
\frac{-\sqrt{\kappa_e} \ain}{\Aa}\\
\frac{-\sqrt{\gamma_e} \cin}{\Ab}
\end{bmatrix}.
\end{equation}
Similarly, the electro-optic $S$ parameter is
\begin{eqnarray}
S_{\textrm{ac}} &=& \sqrt{\kappae} \frac{\etaab}{1-\etaaba} \frac{-\sqrt{\gammae}}{\Ab}\\
&=&\sqrt{\kappae} \frac{1}{\Aa} \frac{iG}{\Ab+ G^{2}/\Aa}\sqrt{\gammae}
\end{eqnarray}
The resulting conversion efficiency at perfectly-matched frequency $\omega = \Delta = \omegam$ is
\begin{eqnarray}
\eta &= & \kappae \frac{4}{\kappa^{2}} \frac{4G^{2}/\gamma^{2} }{ (1 + 4G^{2}/\kappa\gamma)^{2} } \gammae\\
& = & \eta_{\textrm{ext, a}} \eta_{\textrm{ext, b}} \frac{4 \Cab}{(1+\Cab)^{2}}
\end{eqnarray}
where $\eta_{\textrm{ext, b}} = \gammae/\gamma $. The maximum of conversion efficiency occurs at the matching condition $\Cab = 1$, where similarly $ \eta =\eta_{\textrm{ext, a}} \eta_{\textrm{ext, b}}  $.

\subsection{Measurement of coupling between the mechanical resonator and a microwave channel}

Our OMC can be coupled to a microwave channel via the piezoelectric interaction. We evaporated electrode on both ends of the nanobeam and connected them to a transmission line with impedance $Z_{0}=50~\Omega$. The electric field generated by the electrode is parallel to the nanobeam. The same configuration can also be adopted for coupling the mechanical resonator to microwave circuits.

From the last section, with $\Delta = \omegam$, the coherent mechanical field amplitude is
\begin{equation}
\beta = \frac{-\sqrt{\gamma_{e}} \cin}{i(\omegam-\omega_{\mu}) + \gammatot/2},
\end{equation}
where $\cin$ is the input microwave amplitude with a unit of $1/\sqrt{\textrm{Hz}}$, $\gammatot=  \gammai + \gammae + \gammaOM$ is the total mechanical linewidth and $\omega_{\mu} \approx \omegam$ is the microwave signal frequency. The resulting coherent phonon number in the mechanical mode is
\begin{equation}
n_{\textrm{coh}} = |\beta|^{2} = \frac{\gamma_{e} \dot{N}_{\mu}}{(\omega_{\mu}-\omega_m)^2 + (\gammatot/2)^2},
\end{equation}
where $\dot{N}_{\mu} = |\cin|^{2} = \eta_{\textrm{loss}} P_{\mu}/(\hbar \omega_{\mu})$ is the input microwave photon flux, $P_{\mu}$ is the output power of the VNA and $\eta_{\textrm{loss}} \approx 58\% $ accounts for external RF cable loss. We tune $\omega_{\mu} = \omegam$ and measure the transduced optical sideband power spectral density of both the coherent phonons and thermal phonons using the RSA. The microwave-to-mechanics coupling rate is given by
\begin{equation}
\gamma_{e} = \frac{n_{\textrm{coh}} \gammatot^{2}}{4\dot{N}_{\mu}}.
\end{equation}

\begin{figure}[h]
	\includegraphics[scale=0.4]{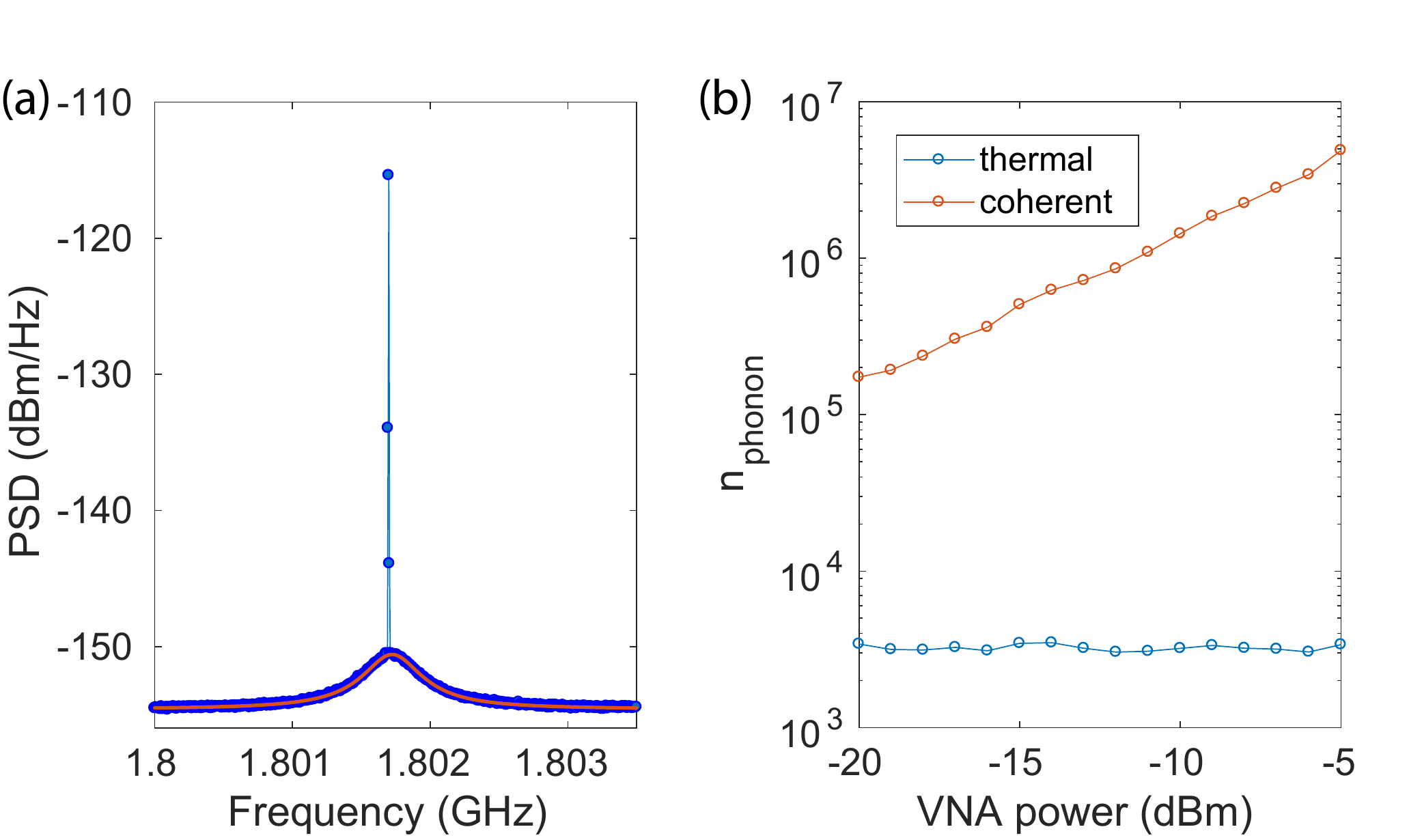}
	\caption{\label{SIFig:MW2Mech_meas} Microwave to mechanics conversion. (a) Measured power spectral density of thermal mechanical motion and coherent mechanical motion from the piezoelectric drive. Blue: data, red: Lorentzian fit of the thermal motion peak. (b) Extracted phonon numbers from the power spectral density for different VNA drive power.}
\end{figure}

Fig.~\ref{SIFig:MW2Mech_meas}(a) shows a measured optical sideband power spectral density (PSD). The thermal mechanical motion give rise to the broad Lorentzian peak and the coherent RF drive corresponds to the sharp peak. We integrate the sideband RF power under both the thermal peak $P_{\textrm{therm}}$ and the coherent peak $P_{\textrm{coh}}$ independently. The thermal occupacy of the mechanical mode is calculated by $ n_{\textrm{therm}} = k_{B}T/\hbar \omegam \approx 3400$ with $T = 295~\textrm{K} $. The optical and electrical gain of the whole readout chain is then determined by $ G = \left<P_{\textrm{therm}}\right>/n_{\textrm{therm}} $ for a certain optical pump power. We drive the OMC with different RF powers and extract the thermal and coherent phonon numbers from the PSD in fig.~\ref{SIFig:MW2Mech_meas}(b). The coherent phonon numbers rise linearly with the increasing RF powers while the thermal phonon occupancies stay near-constant. We obtained $ \gamma_{e}/2\pi = 8.8 \pm 0.56~\textrm{mHz} $.

\subsection{Estimating coupling between the LN OMC and a microwave resonator or a superconducting qubit}

In this section we estimate the coupling of the mechanical mode to a microwave resonator or a superconducting qubit based on our measurement of $\gamma_{e}$ in the last section.

\begin{figure*}[h]
	\centering
	\includegraphics[scale=0.6]{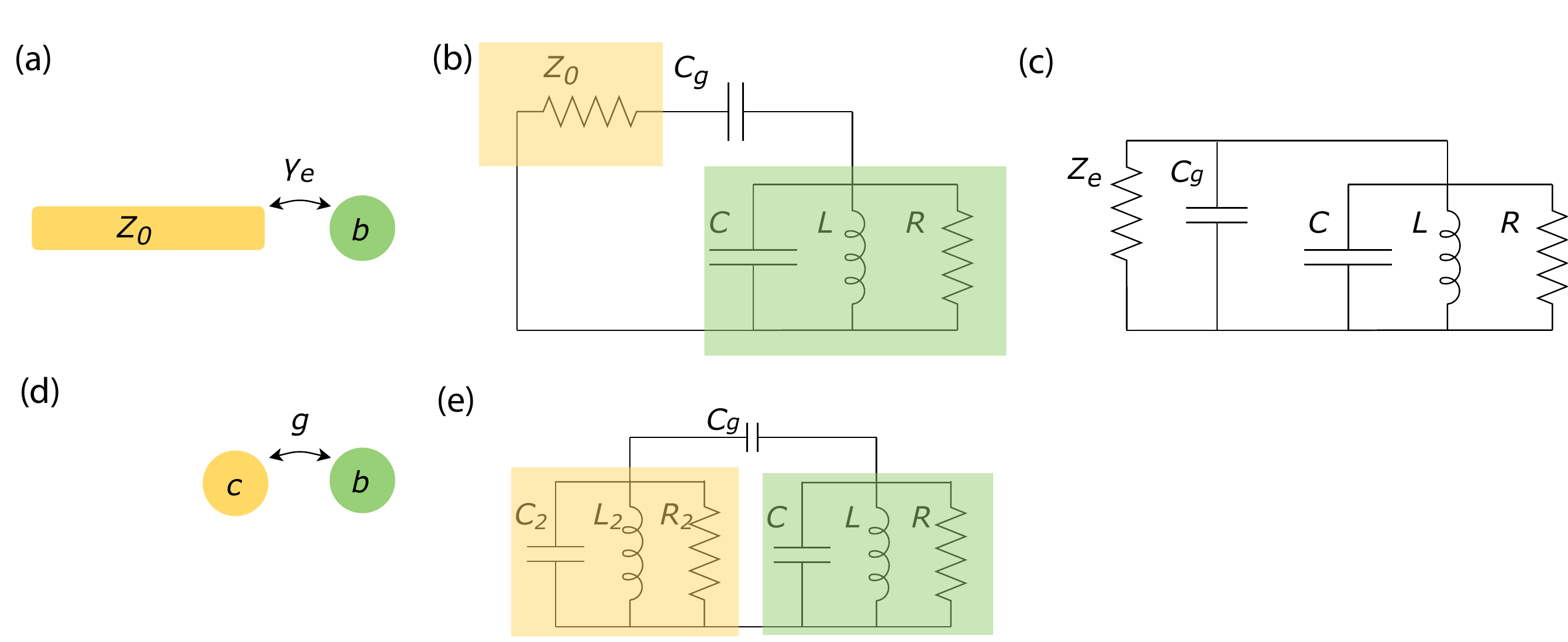}
	\caption{\label{SIFig:mech2MW} Coupling between mechanics and microwave. (a) A mechanical resonator (green) coupled to a microwave channel (yellow). (b) Circuit model of (a). (c) Equivalent circuit of (b). (d) A mechanical resonator (green) coupled to a microwave resonator (yellow). (e) Circuit model of (d). }
\end{figure*}

The mechanical resonator is commonly modeled as  a parallel LC resonator~\cite{Arrangoiz-Arriola2018}. To take into account the non-zero energy decay rate $\gamma$, a resistor is added in parallel such that $\gamma = 1/(RC)$. The mechanical resonator can be coupled to an external $50~\Omega$ transmission line through coupling capacitance $C_{g}$, as shown in fig.~\ref{SIFig:mech2MW}(a). The external coupling introduces a frequency shift and an additional decay rate from an effective conductance $1/Z_{e} = \omegam^{2} C_{g}^{2} Z_{0}$~\cite{schuster2007circuit}. From this equivalence, we obtain a relationship between the coupling capacitance $C_{g}$ and the coupling induced decay rate $\gamma_{e} =  \omegam^{2} C_{g}^{2} Z_{0}/C$.

We show in fig.~\ref{SIFig:mech2MW}(e) a mechanical resonator coupling to a microwave resonator or a superconducting qubit. The coupling rate is given by~\cite{bosman2017approaching}
\begin{eqnarray}
g&\simeq& \sqrt{\omegam\omega_{2}}\cdot \frac{C_g}{2\sqrt{(C+C_{g})(C_{2}+C_{g})}}\\
&\approx&\sqrt{\omegam\omega_{2}}\cdot \frac{C_g}{2\sqrt{CC_{2}}},
\end{eqnarray}
where we've made the assumption that $C_{g}\ll C_{1}, C_{2}$. To simplify the result, we assume that the two modes are perfectly matched with $\omega_{2} = \omegam$, and the microwave resonator has a characteristic impedance $Z_{c} = \sqrt{L_{2}/C_{2}}$. As a result,
\begin{equation}
g^2 = \frac{\omegam^{2} C_{g}^{2}}{C} \cdot \frac{1}{4C_{2}} = \frac{\gamma_{e}}{Z_{0}} \cdot \frac{\omegam Z_{c}}{4},
\end{equation}
where we made the substitution $1/C_{2} = \omega_{2} Z_{c}$.

In conclusion, the coupling rate between a mechanical resonator and a microwave channel $\gamma_{e}$ and the coupling rate between a mechanical resonator and a microwave resonator $g$ are related by $g= \sqrt{\gamma_{e}\omegam} \sqrt{Z_{c}/Z_{0}}/2$.

\section{Extracting pump detunings from on-chip electro-optic modulation}

We start with the electro-optic interaction Hamiltonian. The perturbation on the optical mode frequency from a voltage $V$ across the electodes are modeled by
\begin{eqnarray}
H_{\textrm{EO}} &=& \hbar \frac{d\omega}{dV} V \opd{a}{} \op{a}{}\\
&=& -i\hbar \sqrt{\kappaemu} \opd{a}{} \op{a}{} (\cin - \cin^{\dagger}).
\end{eqnarray}
Where we rewrite the voltage in terms of input microwave amplitude $\cin$ and choose a specific phase of $\cin$ for later convenience. After linearization and rotating wave approximation:
\begin{equation}
H_{\textrm{EO}} = -i \hbar \sqrt{\kappaemu}  ( \alpha_{0} \opd{a}{} \cin  - \alpha_{0}^{*} \op{a}{} \cin^{\dagger}).
\end{equation}
where $\alpha_{0} = -\sqrt{\kappae} \alpha_{\textrm{in}} /(i\Delta + \kappa/2) $ is the optical intracavity pump amplitude and $\alpha_{\textrm{in}}$ is the pump input photon amplitude. Equation of motion for the optical sideband amplitude is
\begin{equation}
- i\omega a(\omega) = -(i \Delta + \frac{\kappa}{2}) a - \sqrt{\kappaemu } \alpha_{0}  \cin
\end{equation}
from which we could solve the sideband amplitude and calculate the output optical field as
\begin{eqnarray}
\alpha_{\textrm{out}}& =& \left[ 1-\frac{\kappae}{i \Delta + \frac{\kappa}{2}} \left( 1- \frac{\sqrt{\kappaemu}}{i(\Delta - \omegamu) + \frac{\kappa}{2}} \cin e^{-i\omegamu t} \right) \right] \alpha_{\textrm{in}}\notag\\
&=& \left( r(\Delta) + s_{\textrm{EO}}(\omegamu, \Delta) \cin e^{-i\omegamu t} \right)\alpha_{\textrm{in}},
\end{eqnarray}
where
\begin{equation}
r(\Delta) = 1 - \frac{\kappae}{i\Delta + \kappa/2},
\end{equation}
\begin{equation}
s_{\textrm{EO}}(\omega, \Delta)  = \frac{\kappae}{i\Delta + \kappa/2}\frac{\sqrt{\kappaemu}}{i(\Delta - \omega) + \frac{\kappa}{2}}.
\end{equation}

The electronic signal from the high-speed photo-detector is
\begin{eqnarray}
V_{\textrm{HS}} &=& G |\alpha_{\textrm{out}}|^{2} =G |\alpha_{\textrm{in}}|^{2} \left( |r(\Delta)|^{2} +\right. \notag \\
&& \left. r^{*}(\Delta) s_{\textrm{EO}} (\omegamu, \Delta)\cin e^{-i\omegamu t} + h.c.  + O(\cin^{2})\right)
\end{eqnarray}
where $G$ denotes the total detection gain. As a result, the VNA directly measures $S_{21}(\omega, \Delta) \propto G|\alpha_{\textrm{in}}|^{2}r^{*}(\Delta) s_{\textrm{EO}} (\omega, \Delta) S_{\textrm{ext}} (\omega) $ where we further include the external response of the cables and wirebonds in $S_{\textrm{ext}} (\omega)$. 

For the coherent spectroscopy, the external response and the detection gain can be removed by dividing the response with a background taken with far-detuned pump laser~\cite{chan2012thesis,patel2018single}. The electro-optic (EO) sideband is generated by the EOM and doesn't depend on the cavity-laser detuning. However, for the on-chip EO modulation, the intracavity photon numbers are much smaller for far-detuned pump laser, resulting in a much weaker EO sideband and a very low SNR. To eliminate the detection gain $G$ and the external response, we take the ratio between two $S_{21}$ measurements with different near-cavity detunings $\Delta_{1}$ and $\Delta_{2}$ where $|\Delta_{1}-\Delta_{2}| \gtrsim \kappa$. The gain and external response is identical and cancels, leaving only the ratio of the on-chip EO response. We denote the ratio of the $S_{21}$ as
\begin{equation}
\label{SIEQ:OnChipEOMResp}
S_{\Delta_{2}/\Delta_{1}} (\omega) = \frac{r^{*}(\Delta_{2}) s_{\textrm{EO}} (\omega, \Delta_{2})}{r^{*}(\Delta_{1}) s_{\textrm{EO}} (\omega, \Delta_{1})}.
\end{equation}
By fitting the measured $S_{21}$ ratio with different detunings and identical pump laser power, we extract both detunings $\Delta_{1} $ and $\Delta_{2}$.

\begin{figure}[h]
	\includegraphics[scale=0.45]{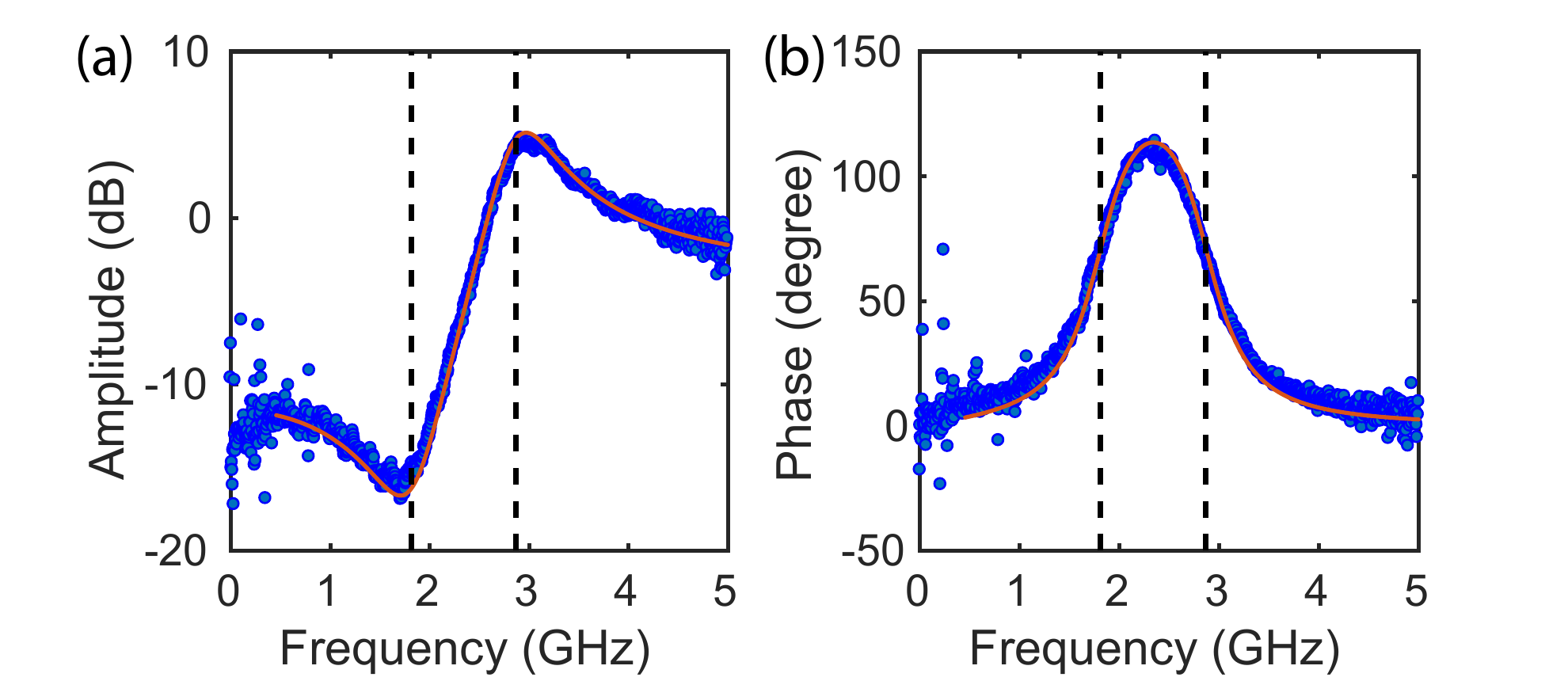}
	\caption{\label{SIFig:OnChipEOM} On-chip electro-optic response and fit. (a) Amplitude and (b) phase of the electro-optic response. Red curves show the fitting results. The detunings extracted from the fit are shown as black vertical dashed lines for comparison.}
\end{figure}

In fig.~\ref{SIFig:OnChipEOM} we show a typical measured $S_{\Delta_{2}/\Delta_{1}}$ (blue) and the fitting results using eq.~\ref{SIEQ:OnChipEOMResp} (red). The external responses are perfectly removed, allowing us to extract the detunings of the pump laser. We point out that with the piezoelectric drive off, the pump detunings could also be extracted using the external EOM sideband sweep as used in the coherent spectroscopy.

\onecolumn

\section{Rotated photoelastic tensor components}
\label{SI:RotatedPEComponents}

The original photoelastic tensor components are given by~\cite{Weis1985}
\begin{equation}
p = \begin{bmatrix}
p_{11} & p_{12} & p_{13} & p_{14} & 0 & 0 \\
p_{12} & p_{11} & p_{13} & -p_{14} & 0 & 0 \\
p_{31} & p_{31} & p_{33} & 0 & 0 & 0 \\
p_{41} & -p_{41} & 0 & p_{44} & 0 & 0 \\
0 & 0 & 0 & 0 & p_{44} & p_{41} \\
0 & 0 & 0 & 0 & p_{14} & (p_{11}-p_{12})/2
\end{bmatrix}.
\end{equation}
Using the rotation matrix and MATHEMATICA~\cite{MMA}, we obtain the rotated photoelastic components in the contracted index notation for X-cut LN as
\begin{equation}
p'_{\textrm{LNX}} = \begin{bmatrix}
p'_{11} & p'_{12} & p'_{13} & 0 & 0 & p'_{16} \\
p'_{21} & p'_{22} & p'_{23} & 0 & 0 & p'_{26} \\
p'_{31} & p'_{32} & p'_{33} & 0 & 0 & p'_{36} \\
0 & 0 & 0 & p'_{44} & p'_{45} & 0 \\
0 & 0 & 0 & p'_{54} & p'_{55} & 0 \\
p'_{61} & p'_{62} & p'_{63} & 0 & 0 & p'_{66}
\end{bmatrix},
\end{equation}
with
\begin{eqnarray}
p'_{11} &=& p_{11} \sin ^4(\phi )+p_{44} \sin ^2(2 \phi )+p_{33} \cos ^4(\phi )  +p_{13} \sin
^2(\phi ) \cos ^2(\phi )+p_{31} \sin ^2(\phi ) \cos ^2(\phi ) \notag \\
& & -2 p_{14} \sin
^3(\phi ) \cos (\phi )-2 p_{41} \sin ^3(\phi ) \cos (\phi ),\\
p'_{12} &=&p_{13} \sin ^4(\phi )-p_{44} \sin ^2(2 \phi )+p_{31} \cos ^4(\phi )-2 p_{41} \sin
(\phi ) \cos ^3(\phi )+p_{11} \sin ^2(\phi ) \cos ^2(\phi )\notag \\
& & +p_{33} \sin ^2(\phi )
\cos ^2(\phi )+2 p_{14} \sin ^3(\phi ) \cos (\phi )\\
p'_{13} &=& p_{12} \sin ^2(\phi )+p_{41} \sin (2 \phi )+p_{31} \cos ^2(\phi )\\
p'_{16} &=& \sin (\phi ) \left(-p_{31} \cos ^3(\phi )+p_{33} \cos ^3(\phi )-2 p_{44} \cos (2 \phi
) \cos (\phi )+2 p_{41} \sin (\phi ) \cos ^2(\phi )-p_{11} \sin ^2(\phi ) \cos
(\phi ) \right . \notag \\
& & \left. +p_{13} \sin ^2(\phi ) \cos (\phi )+p_{14} \sin (\phi ) \cos (2 \phi    )\right) 
\end{eqnarray}
\begin{eqnarray}
p'_{21} &=&p_{31} \sin ^4(\phi )-p_{44} \sin ^2(2 \phi )+p_{13} \cos ^4(\phi )-2 p_{14} \sin
(\phi ) \cos ^3(\phi )+p_{11} \sin ^2(\phi ) \cos ^2(\phi ) \notag \\
&& +p_{33} \sin ^2(\phi )
\cos ^2(\phi )+2 p_{41} \sin ^3(\phi ) \cos (\phi )\\
p'_{22} & = & p_{33} \sin ^4(\phi )+p_{44} \sin ^2(2 \phi )+p_{11} \cos ^4(\phi )+2 p_{14} \sin
(\phi ) \cos ^3(\phi )+2 p_{41} \sin (\phi ) \cos ^3(\phi ) \notag \\
&& +p_{13} \sin ^2(\phi )
\cos ^2(\phi )+p_{31} \sin ^2(\phi ) \cos ^2(\phi )\\
p'_{23} &=& p_{12} \cos ^2(\phi )+\sin (\phi ) \left(p_{31} \sin (\phi )-2 p_{41} \cos (\phi
)\right)\\
p'_{26} &=& \cos (\phi ) \left(-p_{31} \sin ^3(\phi )+p_{33} \sin ^3(\phi )+p_{14} \cos (\phi )
\cos (2 \phi )-p_{11} \sin (\phi ) \cos ^2(\phi )+p_{13} \sin (\phi ) \cos ^2(\phi
) \right. \notag \\
& & \left. -2 p_{41} \sin ^2(\phi ) \cos (\phi )+2 p_{44} \sin (\phi ) \cos (2 \phi )\right)
\end{eqnarray}
\begin{eqnarray}
p'_{31} &=&p_{12} \sin ^2(\phi )+p_{14} \sin (2 \phi )+p_{13} \cos ^2(\phi )\\
p'_{32} &=&p_{12} \cos ^2(\phi )+\sin (\phi ) \left(p_{13} \sin (\phi )-2 p_{14} \cos (\phi
)\right)\\
p'_{33} &=& p_{11}\\
p'_{36} &=& \frac{1}{2} \left(p_{12} (-\sin (2 \phi ))+p_{13} \sin (2 \phi )-2 p_{14} \cos (2
\phi )\right)
\end{eqnarray}
\begin{eqnarray}
p'_{44} &=& \frac{1}{2} \left(p_{11}-p_{12}\right) \cos ^2(\phi )-p_{14} \sin (\phi ) \cos (\phi
) +\sin (\phi ) \left(p_{44} \sin (\phi )-p_{41} \cos (\phi )\right)\\
p'_{45} &=& \sin (\phi ) \left(p_{41} \sin (\phi )+p_{44} \cos (\phi )\right)-\cos (\phi )
\left(\frac{1}{2} \left(p_{11}-p_{12}\right) \sin (\phi )+p_{14} \cos (\phi
)\right) \\
p'_{54} &=& \sin (\phi ) \left(p_{14} \sin (\phi )-\frac{1}{2} \left(p_{11}-p_{12}\right) \cos
(\phi )\right)+\cos (\phi ) \left(p_{44} \sin (\phi )-p_{41} \cos (\phi )\right) \\
p'_{55} &=& \sin (\phi ) \left(\frac{1}{2} \left(p_{11}-p_{12}\right) \sin (\phi )+p_{14} \cos
(\phi )\right)+\cos (\phi ) \left(p_{41} \sin (\phi )+p_{44} \cos (\phi )\right)
\end{eqnarray}
\begin{eqnarray}
p'_{61} &=& \sin (\phi ) \left(-p_{41} \sin ^3(\phi )+p_{44} \sin (\phi ) \sin (2 \phi )-p_{13}
\cos ^3(\phi )+p_{33} \cos ^3(\phi )-2 p_{44} \cos ^3(\phi )+2 p_{14} \sin (\phi )
\cos ^2(\phi ) \right. \notag \\
& & \left. +p_{41} \sin (\phi ) \cos ^2(\phi ) -p_{11} \sin ^2(\phi ) \cos (\phi   )+p_{31} \sin ^2(\phi ) \cos (\phi )\right)\\
p'_{62} &=& \cos (\phi ) \left(p_{33} \sin ^3(\phi )+p_{31} \sin (\phi ) \cos ^2(\phi )  +\cos ^2(\phi ) \left(2 p_{44} \sin (\phi )+p_{41} \cos (\phi )\right)  \right. \notag \\
& & \left. -\sin   (\phi ) \left(p_{11} \cos ^2(\phi )+\sin (\phi ) \left(p_{13} \sin (\phi )+2  p_{44} \sin (\phi ) +2 p_{14} \cos (\phi )+p_{41} \cos (\phi )\right)\right) \right) \\
p'_{63} &=&\frac{1}{2} \left(p_{12} (-\sin (2 \phi ))+p_{31} \sin (2 \phi )-2 p_{41} \cos (2
\phi )\right)\\
p'_{66} &=& \frac{1}{8} \left(2 p_{11} \sin ^2(2 \phi )-2 p_{14} \sin (4 \phi )-2 p_{41} \sin (4
\phi )+p_{13} (\cos (4 \phi )-1)+p_{31} \cos (4 \phi ) \right. \notag \\
& & \left.-p_{33} \cos (4 \phi )+4
p_{44} \cos (4 \phi )-p_{31}+p_{33}+4 p_{44}\right)
\end{eqnarray}

For Y-cut LN, the rotated photoelastic tensor components are given by
\begin{equation}
p'_{\textrm{LNY}} = \begin{bmatrix}
p'_{11} & p'_{12} & p'_{13} & p'_{14} & p'_{15} & p'_{16} \\
p'_{21} & p'_{22} & p'_{23} & p'_{24} & p'_{25} & p'_{26} \\
p'_{31} & p'_{32} & p'_{33} & p'_{34} & p'_{35} & p'_{36} \\
p'_{41} & p'_{42} & p'_{43} & p'_{44} & p'_{45} & p'_{46} \\
p'_{51} & p'_{52} & p'_{53} & p'_{54} & p'_{55} & p'_{56} \\
p'_{61} & p'_{62} & p'_{63} & p'_{64} & p'_{65} & p'_{66}
\end{bmatrix},
\end{equation}
where
\begin{eqnarray}
p'_{11}&=&p_{11} \sin ^4(\phi )+p_{44} \sin ^2(2 \phi )+p_{33} \cos ^4(\phi )+p_{13} \sin ^2(\phi ) \cos ^2(\phi )+p_{31} \sin ^2(\phi ) \cos ^2(\phi )\\p'_{12}&=&p_{13} \sin ^4(\phi )-p_{44} \sin ^2(2 \phi )+p_{31} \cos ^4(\phi )+p_{11} \sin ^2(\phi ) \cos ^2(\phi )+p_{33} \sin ^2(\phi ) \cos ^2(\phi )\\p'_{13}&=&p_{12} \sin ^2(\phi )+p_{31} \cos ^2(\phi )\\p'_{14}&=&p_{14} \sin ^3(\phi )-2 p_{41} \sin (\phi ) \cos ^2(\phi )\\p'_{15}&=&\left(p_{14}+2 p_{41}\right) \sin ^2(\phi ) \cos (\phi )\\p'_{16}&=&\sin (\phi ) \cos (\phi ) \left(-p_{11} \sin ^2(\phi )+p_{13} \sin ^2(\phi )-p_{31} \cos ^2(\phi )+p_{33} \cos ^2(\phi )-2 p_{44} \cos (2 \phi )\right)
\end{eqnarray}
\begin{eqnarray}
p'_{21}&=&p_{31} \sin ^4(\phi )-p_{44} \sin ^2(2 \phi )+p_{13} \cos ^4(\phi )+p_{11} \sin ^2(\phi ) \cos ^2(\phi )+p_{33} \sin ^2(\phi ) \cos ^2(\phi )\\p'_{22}&=&p_{11} \cos ^4(\phi )+\sin ^2(\phi ) \left(p_{33} \sin ^2(\phi )+p_{13} \cos ^2(\phi )+p_{31} \cos ^2(\phi )+4 p_{44} \cos ^2(\phi )\right)\\p'_{23}&=&p_{31} \sin ^2(\phi )+p_{12} \cos ^2(\phi )\\p'_{24}&=&\left(p_{14}+2 p_{41}\right) \sin (\phi ) \cos ^2(\phi )\\p'_{25}&=&p_{14} \cos ^3(\phi )-2 p_{41} \sin ^2(\phi ) \cos (\phi )\\p'_{26}&=&\sin (\phi ) \cos (\phi ) \left(-p_{31} \sin ^2(\phi )+p_{33} \sin ^2(\phi )-p_{11} \cos ^2(\phi )+p_{13} \cos ^2(\phi )+2 p_{44} \cos (2 \phi )\right)
\end{eqnarray}
\begin{eqnarray}
p'_{31}&=&p_{12} \sin ^2(\phi )+p_{13} \cos ^2(\phi )\\p'_{32}&=&p_{13} \sin ^2(\phi )+p_{12} \cos ^2(\phi )\\p'_{33}&=&p_{11}\\p'_{34}&=&p_{14} (-\sin (\phi ))\\p'_{35}&=&p_{14} (-\cos (\phi ))\\p'_{36}&=&\left(p_{13}-p_{12}\right) \sin (\phi ) \cos (\phi )
\end{eqnarray}
\begin{eqnarray}p'_{41}&=&p_{41} \sin ^3(\phi )-2 p_{14} \sin (\phi ) \cos ^2(\phi )\\p'_{42}&=&\left(2 p_{14}+p_{41}\right) \sin (\phi ) \cos ^2(\phi )\\p'_{43}&=&p_{41} (-\sin (\phi ))\\p'_{44}&=&p_{44} \sin ^2(\phi )+\frac{1}{2} \left(p_{11}-p_{12}\right) \cos ^2(\phi )\\p'_{45}&=&\frac{1}{2} \left(-p_{11}+p_{12}+2 p_{44}\right) \sin (\phi ) \cos (\phi )\\p'_{46}&=&\cos (\phi ) \left(p_{14} \cos (2 \phi )-p_{41} \sin ^2(\phi )\right)
\end{eqnarray}
\begin{eqnarray}p'_{51}&=&\left(2 p_{14}+p_{41}\right) \sin ^2(\phi ) \cos (\phi )\\p'_{52}&=&p_{41} \cos ^3(\phi )-2 p_{14} \sin ^2(\phi ) \cos (\phi )\\p'_{53}&=&p_{41} (-\cos (\phi ))\\p'_{54}&=&\frac{1}{2} \left(-p_{11}+p_{12}+2 p_{44}\right) \sin (\phi ) \cos (\phi )\\p'_{55}&=&\frac{1}{2} \left(p_{11}-p_{12}\right) \sin ^2(\phi )+p_{44} \cos ^2(\phi )\\p'_{56}&=&-\sin (\phi ) \left(p_{41} \cos ^2(\phi )+p_{14} \cos (2 \phi )\right)
\end{eqnarray}
\begin{eqnarray}
p'_{61}&=&\sin (\phi ) \cos (\phi ) \left(-p_{11} \sin ^2(\phi )+p_{31} \sin ^2(\phi )+2 p_{44} \sin ^2(\phi )-p_{13} \cos ^2(\phi )+\left(p_{33}-2 p_{44}\right) \cos ^2(\phi )\right)\\
p'_{62}&=&\cos (\phi ) \left(p_{33} \sin ^3(\phi )+p_{31} \sin (\phi ) \cos ^2(\phi )+p_{44} \sin (2 \phi ) \cos (\phi )\right) \notag \\
& & -\sin (\phi ) \left(p_{11} \cos ^3(\phi )+\left(p_{13}+2 p_{44}\right) \sin ^2(\phi ) \cos (\phi )\right)\\
p'_{63}&=&\left(p_{31}-p_{12}\right) \sin (\phi ) \cos (\phi )\\p'_{64}&=&\cos (\phi ) \left(p_{41} \cos (2 \phi )-p_{14} \sin ^2(\phi )\right)\\p'_{65}&=&-\sin (\phi ) \left(p_{14} \cos ^2(\phi )+p_{41} \cos (2 \phi )\right)\\p'_{66}&=&\cos ^2(\phi ) \left(\left(p_{33}-p_{31}\right) \sin ^2(\phi )+p_{44} \cos (2 \phi )\right)-\sin ^2(\phi ) \left(\left(p_{13}-p_{11}\right) \cos ^2(\phi )+p_{44} \cos (2 \phi )\right)
\end{eqnarray}


\begin{thebibliography}{10}
	\newcommand{\enquote}[1]{``#1''}
	
	\bibitem{chan2011laser}
	J.~Chan, T.~M. Alegre, A.~H. Safavi-Naeini, J.~T. Hill, A.~Krause,
	S.~Gr{\"o}blacher, M.~Aspelmeyer, and O.~Painter, \enquote{Laser cooling of a
		nanomechanical oscillator into its quantum ground state,}
	{\protect\JournalTitle{Nature}} \textbf{478}, 89 (2011).
	
	\bibitem{PhysRevLett.108.033602}
	A.~H. Safavi-Naeini, J.~Chan, J.~T. Hill, T.~P.~M. Alegre, A.~Krause, and
	O.~Painter, \enquote{Observation of quantum motion of a nanomechanical
		resonator,} {\protect\JournalTitle{Phys. Rev. Lett.}} \textbf{108}, 033602
	(2012).
	
	\bibitem{hill2012coherent}
	J.~T. Hill, A.~H. Safavi-Naeini, J.~Chan, and O.~Painter, \enquote{Coherent
		optical wavelength conversion via cavity optomechanics,}
	{\protect\JournalTitle{Nature Communications}} \textbf{3}, 1196 (2012).
	
	\bibitem{cohen2015phonon}
	J.~D. Cohen, S.~M. Meenehan, G.~S. MacCabe, S.~Gr{\"o}blacher, A.~H.
	Safavi-Naeini, F.~Marsili, M.~D. Shaw, and O.~Painter, \enquote{Phonon
		counting and intensity interferometry of a nanomechanical resonator,}
	{\protect\JournalTitle{Nature}} \textbf{520}, 522 (2015).
	
	\bibitem{marinkovic2018optomechanical}
	I.~Marinkovi{\'c}, A.~Wallucks, R.~Riedinger, S.~Hong, M.~Aspelmeyer, and
	S.~Gr{\"o}blacher, \enquote{Optomechanical bell test,}
	{\protect\JournalTitle{Physical Review Letters}} \textbf{121}, 220404 (2018).
	
	\bibitem{PhysRevLett.105.220501}
	K.~Stannigel, P.~Rabl, A.~S. S\o{}rensen, P.~Zoller, and M.~D. Lukin,
	\enquote{Optomechanical transducers for long-distance quantum communication,}
	{\protect\JournalTitle{Phys. Rev. Lett.}} \textbf{105}, 220501 (2010).
	
	\bibitem{fang2016optical}
	K.~Fang, M.~H. Matheny, X.~Luan, and O.~Painter, \enquote{Optical transduction
		and routing of microwave phonons in cavity-optomechanical circuits,}
	{\protect\JournalTitle{Nature Photonics}} \textbf{10}, 489 (2016).
	
	\bibitem{fang2017generalized}
	K.~Fang, J.~Luo, A.~Metelmann, M.~H. Matheny, F.~Marquardt, A.~A. Clerk, and
	O.~Painter, \enquote{Generalized non-reciprocity in an optomechanical circuit
		via synthetic magnetism and reservoir engineering,}
	{\protect\JournalTitle{Nature Physics}} \textbf{13}, 465 (2017).
	
	\bibitem{patel2018single}
	R.~N. Patel, Z.~Wang, W.~Jiang, C.~J. Sarabalis, J.~T. Hill, and A.~H.
	Safavi-Naeini, \enquote{Single-mode phononic wire,}
	{\protect\JournalTitle{Physical Review Letters}} \textbf{121}, 040501 (2018).
	
	\bibitem{riedinger2018remote}
	R.~Riedinger, A.~Wallucks, I.~Marinkovi{\'c}, C.~L{\"o}schnauer, M.~Aspelmeyer,
	S.~Hong, and S.~Gr{\"o}blacher, \enquote{Remote quantum entanglement between
		two micromechanical oscillators,} {\protect\JournalTitle{Nature}}
	\textbf{556}, 473 (2018).
	
	\bibitem{chan2012optimized}
	J.~Chan, A.~H. Safavi-Naeini, J.~T. Hill, S.~Meenehan, and O.~Painter,
	\enquote{Optimized optomechanical crystal cavity with acoustic radiation
		shield,} {\protect\JournalTitle{Applied Physics Letters}} \textbf{101},
	081115 (2012).
	
	\bibitem{Weinstein2010}
	D.~Weinstein and S.~A. Bhave, \enquote{{The Resonant Body Transistor},}
	{\protect\JournalTitle{Nano Letters}} \textbf{10}, 1234--1237 (2010).
	
	\bibitem{VanLaer2018}
	R.~{Van Laer}, R.~N. Patel, T.~P. McKenna, J.~D. Witmer, and A.~H.
	Safavi-Naeini, \enquote{{Electrical driving of X-band mechanical waves in a
			silicon photonic circuit},} {\protect\JournalTitle{APL Photonics}}
	\textbf{3}, 086102 (2018).
	
	\bibitem{Kalaee2019}
	M.~Kalaee, M.~Mirhosseini, P.~B. Dieterle, M.~Peruzzo, J.~M. Fink, and
	O.~Painter, \enquote{{Quantum electromechanics of a hypersonic crystal},}
	{\protect\JournalTitle{Nature Nanotechnology}}  (2019).
	
	\bibitem{pernice2012high}
	W.~Pernice, C.~Xiong, C.~Schuck, and H.~Tang, \enquote{High-q aluminum nitride
		photonic crystal nanobeam cavities,} {\protect\JournalTitle{Applied Physics
			Letters}} \textbf{100}, 091105 (2012).
	
	\bibitem{fan2013aluminum}
	L.~Fan, X.~Sun, C.~Xiong, C.~Schuck, and H.~X. Tang, \enquote{Aluminum nitride
		piezo-acousto-photonic crystal nanocavity with high quality factors,}
	{\protect\JournalTitle{Applied Physics Letters}} \textbf{102}, 153507 (2013).
	
	\bibitem{bochmann2013nanomechanical}
	J.~Bochmann, A.~Vainsencher, D.~D. Awschalom, and A.~N. Cleland,
	\enquote{Nanomechanical coupling between microwave and optical photons,}
	{\protect\JournalTitle{Nature Physics}} \textbf{9}, 712 (2013).
	
	\bibitem{vainsencher2016bi}
	A.~Vainsencher, K.~Satzinger, G.~Peairs, and A.~Cleland,
	\enquote{Bi-directional conversion between microwave and optical frequencies
		in a piezoelectric optomechanical device,} {\protect\JournalTitle{Applied
			Physics Letters}} \textbf{109}, 033107 (2016).
	
	\bibitem{ding2011wavelength}
	L.~Ding, C.~Baker, P.~Senellart, A.~Lemaitre, S.~Ducci, G.~Leo, and I.~Favero,
	\enquote{Wavelength-sized gaas optomechanical resonators with gigahertz
		frequency,} {\protect\JournalTitle{Applied Physics Letters}} \textbf{98},
	113108 (2011).
	
	\bibitem{balram2014moving}
	K.~C. Balram, M.~Davan{\c{c}}o, J.~Y. Lim, J.~D. Song, and K.~Srinivasan,
	\enquote{Moving boundary and photoelastic coupling in gaas optomechanical
		resonators,} {\protect\JournalTitle{Optica}} \textbf{1}, 414--420 (2014).
	
	\bibitem{balram2016coherent}
	K.~C. Balram, M.~I. Davan{\c{c}}o, J.~D. Song, and K.~Srinivasan,
	\enquote{Coherent coupling between radiofrequency, optical and acoustic waves
		in piezo-optomechanical circuits,} {\protect\JournalTitle{Nature photonics}}
	\textbf{10}, 346 (2016).
	
	\bibitem{forsch2018microwave}
	M.~Forsch, R.~Stockill, A.~Wallucks, I.~Marinkovic, C.~G{\"a}rtner, R.~A.
	Norte, F.~van Otten, A.~Fiore, K.~Srinivasan, and S.~Gr{\"o}blacher,
	\enquote{Microwave-to-optics conversion using a mechanical oscillator in its
		quantum groundstate,} {\protect\JournalTitle{arXiv preprint
			arXiv:1812.07588}}  (2018).
	
	\bibitem{mitchell2014cavity}
	M.~Mitchell, A.~C. Hryciw, and P.~E. Barclay, \enquote{Cavity optomechanics in
		gallium phosphide microdisks,} {\protect\JournalTitle{Applied Physics
			Letters}} \textbf{104}, 141104 (2014).
	
	\bibitem{schneider2017optomechanics}
	K.~Schneider, P.~Welter, Y.~Baumgartner, S.~H{\"o}nl, H.~Hahn, L.~Czornomaz,
	and P.~Seidler, \enquote{Optomechanics with one-dimensional gallium phosphide
		photonic crystal cavities,} in \emph{Quantum Nanophotonics,} , vol. 10359
	(International Society for Optics and Photonics, 2017), p. 103590K.
	
	\bibitem{Arrangoiz-Arriola2018}
	P.~Arrangoiz-Arriola, E.~A. Wollack, M.~Pechal, J.~D. Witmer, J.~T. Hill, and
	A.~H. Safavi-Naeini, \enquote{Coupling a superconducting quantum circuit to a
		phononic crystal defect cavity,} {\protect\JournalTitle{Phys. Rev. X}}
	\textbf{8}, 031007 (2018).
	
	\bibitem{shao2019high}
	L.~Shao, S.~Maity, L.~Wu, A.~Shams-Ansari, Y.-I. Sohn, E.~Puma, M.~Gadalla,
	M.~Zhang, C.~Wang, and M.~Lon{\v{c}}ar, \enquote{High-q gigahertz surface
		acoustic wave cavity on lithium niobate,} {\protect\JournalTitle{arXiv
			preprint arXiv:1901.09080}}  (2019).
	
	\bibitem{Liang2017}
	H.~Liang, R.~Luo, Y.~He, H.~Jiang, and Q.~Lin, \enquote{{High-quality Lithium
			Niobate Photonic Crystal Nanocavities},} {\protect\JournalTitle{Optica}}
	\textbf{4}, 1251 (2017).
	
	\bibitem{li2018high}
	M.~Li, H.~Liang, R.~Luo, Y.~He, and Q.~Lin, \enquote{High-q two-dimensional
		lithium niobate photonic crystal slab nanoresonators,}
	{\protect\JournalTitle{arXiv preprint arXiv:1806.04755}}  (2018).
	
	\bibitem{wang2018ultrahigh}
	C.~Wang, C.~Langrock, A.~Marandi, M.~Jankowski, M.~Zhang, B.~Desiatov, M.~M.
	Fejer, and M.~Lon{\v{c}}ar, \enquote{Ultrahigh-efficiency wavelength
		conversion in nanophotonic periodically poled lithium niobate waveguides,}
	{\protect\JournalTitle{Optica}} \textbf{5}, 1438--1441 (2018).
	
	\bibitem{safavi2011proposal}
	A.~H. Safavi-Naeini and O.~Painter, \enquote{Proposal for an optomechanical
		traveling wave phonon--photon translator,} {\protect\JournalTitle{New Journal
			of Physics}} \textbf{13}, 013017 (2011).
	
	\bibitem{Aspelmeyer2014}
	M.~Aspelmeyer, T.~J. Kippenberg, and F.~Marquardt, \enquote{{Cavity
			optomechanics},} {\protect\JournalTitle{Revs. Mod. Phys.}} \textbf{86},
	1391--1452 (2014).
	
	\bibitem{Safavi-Naeini2019}
	A.~H. Safavi-Naeini, D.~{Van Thourhout}, R.~Baets, and R.~{Van Laer},
	\enquote{{Controlling phonons and photons at the wavelength scale: integrated
			photonics meets integrated phononics},} {\protect\JournalTitle{Optica}}
	\textbf{6}, 213 (2019).
	
	\bibitem{kippenberg2005analysis}
	T.~Kippenberg, H.~Rokhsari, T.~Carmon, A.~Scherer, and K.~Vahala,
	\enquote{Analysis of radiation-pressure induced mechanical oscillation of an
		optical microcavity,} {\protect\JournalTitle{Physical Review Letters}}
	\textbf{95}, 033901 (2005).
	
	\bibitem{jiang2016chip}
	W.~C. Jiang and Q.~Lin, \enquote{Chip-scale cavity optomechanics in lithium
		niobate,} {\protect\JournalTitle{Scientific Reports}} \textbf{6}, 36920
	(2016).
	
	\bibitem{o2010quantum}
	A.~D. O’Connell, M.~Hofheinz, M.~Ansmann, R.~C. Bialczak, M.~Lenander,
	E.~Lucero, M.~Neeley, D.~Sank, H.~Wang, M.~Weides \emph{et~al.},
	\enquote{Quantum ground state and single-phonon control of a mechanical
		resonator,} {\protect\JournalTitle{Nature}} \textbf{464}, 697 (2010).
	
	\bibitem{Sletten2019resolving}
	L.~R. Sletten, B.~A. Moores, J.~J. Viennot, and K.~W. Lehnert,
	\enquote{Resolving phonon fock states in a multimode cavity with a
		double-slit qubit,} {\protect\JournalTitle{arXiv preprint arXiv:1902.06344}}
	(2019).
	
	\bibitem{arrangoiz2019resolving}
	P.~Arrangoiz-Arriola, E.~A. Wollack, Z.~Wang, M.~Pechal, W.~Jiang, T.~P.
	McKenna, J.~D. Witmer, and A.~H. Safavi-Naeini, \enquote{Resolving the energy
		levels of a nanomechanical oscillator,} {\protect\JournalTitle{arXiv preprint
			arXiv:1902.04681}}  (2019).
	
	\bibitem{witmer2017high}
	J.~D. Witmer, J.~A. Valery, P.~Arrangoiz-Arriola, C.~J. Sarabalis, J.~T. Hill,
	and A.~H. Safavi-Naeini, \enquote{High-q photonic resonators and
		electro-optic coupling using silicon-on-lithium-niobate,}
	{\protect\JournalTitle{Scientific Reports}} \textbf{7}, 46313 (2017).
	
	\bibitem{wang2018nanophotonic}
	C.~Wang, M.~Zhang, B.~Stern, M.~Lipson, and M.~Lon{\v{c}}ar,
	\enquote{Nanophotonic lithium niobate electro-optic modulators,}
	{\protect\JournalTitle{Optics Express}} \textbf{26}, 1547--1555 (2018).
	
	\bibitem{wang2018integrated}
	C.~Wang, M.~Zhang, X.~Chen, M.~Bertrand, A.~Shams-Ansari, S.~Chandrasekhar,
	P.~Winzer, and M.~Lon{\v{c}}ar, \enquote{Integrated lithium niobate
		electro-optic modulators operating at cmos-compatible voltages,}
	{\protect\JournalTitle{Nature}} \textbf{562}, 101 (2018).
	
	\bibitem{arrangoiz2016engineering}
	P.~Arrangoiz-Arriola and A.~H. Safavi-Naeini, \enquote{Engineering interactions
		between superconducting qubits and phononic nanostructures,}
	{\protect\JournalTitle{Physical Review A}} \textbf{94}, 063864 (2016).
	
	\bibitem{wang2017second}
	C.~Wang, X.~Xiong, N.~Andrade, V.~Venkataraman, X.-F. Ren, G.-C. Guo, and
	M.~Lon{\v{c}}ar, \enquote{Second harmonic generation in nano-structured
		thin-film lithium niobate waveguides,} {\protect\JournalTitle{Optics
			Express}} \textbf{25}, 6963--6973 (2017).
	
	\bibitem{jiang2018nonlinear}
	H.~Jiang, H.~Liang, R.~Luo, X.~Chen, Y.~Chen, and Q.~Lin, \enquote{Nonlinear
		frequency conversion in one dimensional lithium niobate photonic crystal
		nanocavities,} {\protect\JournalTitle{Applied Physics Letters}} \textbf{113},
	021104 (2018).
	
	\bibitem{COMSOL}
	\enquote{Comsol multiphysics 5.0,} \url{https://www.comsol.com}.
	
	\bibitem{patel2017engineering}
	R.~N. Patel, C.~J. Sarabalis, W.~Jiang, J.~T. Hill, and A.~H. Safavi-Naeini,
	\enquote{Engineering phonon leakage in nanomechanical resonators,}
	{\protect\JournalTitle{Physical Review Applied}} \textbf{8}, 041001 (2017).
	
	\bibitem{safavi2010design}
	A.~H. Safavi-Naeini and O.~Painter, \enquote{Design of optomechanical cavities
		and waveguides on a simultaneous bandgap phononic-photonic crystal slab,}
	{\protect\JournalTitle{Optics Express}} \textbf{18}, 14926--14943 (2010).
	
	\bibitem{hu2009lithium}
	H.~Hu, R.~Ricken, and W.~Sohler, \enquote{Lithium niobate photonic wires,}
	{\protect\JournalTitle{Optics express}} \textbf{17}, 24261--24268 (2009).
	
	\bibitem{poberaj2012lithium}
	G.~Poberaj, H.~Hu, W.~Sohler, and P.~Guenter, \enquote{Lithium niobate on
		insulator (lnoi) for micro-photonic devices,} {\protect\JournalTitle{Laser \&
			photonics reviews}} \textbf{6}, 488--503 (2012).
	
	\bibitem{wang2014integrated}
	C.~Wang, M.~J. Burek, Z.~Lin, H.~A. Atikian, V.~Venkataraman, I.-C. Huang,
	P.~Stark, and M.~Lon{\v{c}}ar, \enquote{Integrated high quality factor
		lithium niobate microdisk resonators,} {\protect\JournalTitle{Optics
			express}} \textbf{22}, 30924--30933 (2014).
	
	\bibitem{hartung2008fabrication}
	H.~Hartung, E.-B. Kley, A.~T{\"u}nnermann, T.~Gischkat, F.~Schrempel, and
	W.~Wesch, \enquote{Fabrication of ridge waveguides in zinc-substituted
		lithium niobate by means of ion-beam enhanced etching,}
	{\protect\JournalTitle{Optics letters}} \textbf{33}, 2320--2322 (2008).
	
	\bibitem{zhang2017monolithic}
	M.~Zhang, C.~Wang, R.~Cheng, A.~Shams-Ansari, and M.~Lon{\v{c}}ar,
	\enquote{Monolithic ultra-high-q lithium niobate microring resonator,}
	{\protect\JournalTitle{Optica}} \textbf{4}, 1536--1537 (2017).
	
	\bibitem{krasnokutska2018ultra}
	I.~Krasnokutska, J.-L.~J. Tambasco, X.~Li, and A.~Peruzzo, \enquote{Ultra-low
		loss photonic circuits in lithium niobate on insulator,}
	{\protect\JournalTitle{Optics express}} \textbf{26}, 897--904 (2018).
	
	\bibitem{furukawa2001green}
	Y.~Furukawa, K.~Kitamura, A.~Alexandrovski, R.~Route, M.~Fejer, and G.~Foulon,
	\enquote{Green-induced infrared absorption in mgo doped linbo 3,}
	{\protect\JournalTitle{Applied Physics Letters}} \textbf{78}, 1970--1972
	(2001).
	
	\bibitem{weis2010optomechanically}
	S.~Weis, R.~Rivi{\`e}re, S.~Del{\'e}glise, E.~Gavartin, O.~Arcizet,
	A.~Schliesser, and T.~J. Kippenberg, \enquote{Optomechanically induced
		transparency,} {\protect\JournalTitle{Science}} \textbf{330}, 1520--1523
	(2010).
	
	\bibitem{safavi2011electromagnetically}
	A.~H. Safavi-Naeini, T.~M. Alegre, J.~Chan, M.~Eichenfield, M.~Winger, Q.~Lin,
	J.~T. Hill, D.~E. Chang, and O.~Painter, \enquote{Electromagnetically induced
		transparency and slow light with optomechanics,}
	{\protect\JournalTitle{Nature}} \textbf{472}, 69 (2011).
	
	\bibitem{massel2012multimode}
	F.~Massel, S.~U. Cho, J.-M. Pirkkalainen, P.~J. Hakonen, T.~T. Heikkil{\"a},
	and M.~A. Sillanp{\"a}{\"a}, \enquote{Multimode circuit optomechanics near
		the quantum limit,} {\protect\JournalTitle{Nature Communications}}
	\textbf{3}, 987 (2012).
	
	\bibitem{safavi2014two}
	A.~H. Safavi-Naeini, J.~T. Hill, S.~Meenehan, J.~Chan, S.~Gr{\"o}blacher, and
	O.~Painter, \enquote{Two-dimensional phononic-photonic band gap
		optomechanical crystal cavity,} {\protect\JournalTitle{Physical Review
			Letters}} \textbf{112}, 153603 (2014).
	
	\bibitem{samkharadze2016high}
	N.~Samkharadze, A.~Bruno, P.~Scarlino, G.~Zheng, D.~DiVincenzo, L.~DiCarlo, and
	L.~Vandersypen, \enquote{High-kinetic-inductance superconducting nanowire
		resonators for circuit qed in a magnetic field,}
	{\protect\JournalTitle{Physical Review Applied}} \textbf{5}, 044004 (2016).
	
	\bibitem{barzanjeh2017mechanical}
	S.~Barzanjeh, M.~Wulf, M.~Peruzzo, M.~Kalaee, P.~Dieterle, O.~Painter, and
	J.~Fink, \enquote{Mechanical on-chip microwave circulator,}
	{\protect\JournalTitle{Nature Communications}} \textbf{8}, 953 (2017).
	
	\bibitem{andrushchak2009complete}
	A.~Andrushchak, B.~Mytsyk, H.~Laba, O.~Yurkevych, I.~Solskii, A.~Kityk, and
	B.~Sahraoui, \enquote{Complete sets of elastic constants and photoelastic
		coefficients of pure and mgo-doped lithium niobate crystals at room
		temperature,} {\protect\JournalTitle{Journal of Applied Physics}}
	\textbf{106}, 073510 (2009).
	
	\bibitem{carmon2004dynamical}
	T.~Carmon, L.~Yang, and K.~J. Vahala, \enquote{Dynamical thermal behavior and
		thermal self-stability of microcavities,} {\protect\JournalTitle{Optics
			Express}} \textbf{12}, 4742--4750 (2004).
	
	\bibitem{sun2017nonlinear}
	X.~Sun, H.~Liang, R.~Luo, W.~C. Jiang, X.-C. Zhang, and Q.~Lin,
	\enquote{Nonlinear optical oscillation dynamics in high-q lithium niobate
		microresonators,} {\protect\JournalTitle{Optics Express}} \textbf{25},
	13504--13516 (2017).
	
	\bibitem{schuster2007circuit}
	D.~I. Schuster, \emph{Circuit quantum electrodynamics} (Yale University, 2007).
	
	\bibitem{bosman2017approaching}
	S.~J. Bosman, M.~F. Gely, V.~Singh, D.~Bothner, A.~Castellanos-Gomez, and G.~A.
	Steele, \enquote{Approaching ultrastrong coupling in transmon circuit qed
		using a high-impedance resonator,} {\protect\JournalTitle{Physical Review B}}
	\textbf{95}, 224515 (2017).
	
	\bibitem{chan2012thesis}
	J.~Chan, \enquote{Laser cooling of an optomechanical crystal resonator to its
		quantum ground state of motion,} Ph.D. thesis, California Institute of
	Technology (2012).
	
	\bibitem{Weis1985}
	R.~Weis and T.~Gaylord, \enquote{Lithium niobate: Summary of physical
		properties and crystal structure,} {\protect\JournalTitle{Appl. Phys. A}}
	\textbf{37}, 191--203 (1985).
	
	\bibitem{MMA}
	\enquote{Mathematica,} Available from \url{http://www.wolfram.com}.
	
\end{thebibliography}
\end{document}